\documentclass{article}

\usepackage{arxiv}

\usepackage[utf8]{inputenc}
\usepackage[T1]{fontenc}
\usepackage[hypertexnames=false]{hyperref}
\usepackage{url}
\usepackage{booktabs}
\usepackage{amsfonts} 
\usepackage{nicefrac}
\usepackage{microtype}
\usepackage{amsmath}
\usepackage{graphicx}
\usepackage{multirow}
\usepackage{rotating}
\usepackage{multirow}
\usepackage[table,xcdraw]{xcolor}
\usepackage[nameinlink]{cleveref}

\DeclareMathOperator\erf{erf}

\title{Voxelwise principal component analysis of dynamic [S-methyl-\textsuperscript{11}C]methionine PET data in glioma patients}

\author{
  Corentin~Martens\\
  Department of Nuclear Medicine\\
  H\^opital Erasme, Université libre de Bruxelles\\
  Brussels, Belgium\\
  \texttt{corentin.martens@ulb.ac.be}\\
  \And
  Olivier~Debeir\\
  Laboratory of Image Synthesis and Analysis\\
  Universit\'e libre de Bruxelles\\
  Brussels, Belgium\\
  \And
  Christine~Decaestecker\\
  Laboratory of Image Synthesis and Analysis\\
  Universit\'e libre de Bruxelles\\
  Brussels, Belgium\\
  \And
  Thierry~Metens\\
  Department of Radiology\\
  H\^opital Erasme, Université libre de Bruxelles\\
  Brussels, Belgium\\
  \And
  Laetitia~Lebrun\\
  Department of Pathology\\
  H\^opital Erasme, Université libre de Bruxelles\\
  Brussels, Belgium\\
  \And
  Gil~Leurquin-Sterk\\
  Department of Nuclear Medicine\\
  H\^opital Erasme, Université libre de Bruxelles\\
  Brussels, Belgium\\
  \And
  Nicola~Trotta\\
  Department of Nuclear Medicine\\
  H\^opital Erasme, Université libre de Bruxelles\\
  Brussels, Belgium\\
  \And
  Serge~Goldman\\
  Department of Nuclear Medicine\\
  H\^opital Erasme, Université libre de Bruxelles\\
  Brussels, Belgium\\
  \And
  Gaetan~Van Simaeys\\
  Department of Nuclear Medicine\\
  H\^opital Erasme, Université libre de Bruxelles\\
  Brussels, Belgium\\
}

\begin{document}
\maketitle

\begin{abstract}
Recent works have demonstrated the added value of dynamic amino acid positron emission tomography (PET) for glioma grading and genotyping, biopsy targeting, and recurrence diagnosis. However, most of these studies are exclusively based on hand-crafted qualitative or semi-quantitative dynamic features extracted from the mean time activity curve (TAC) within predefined volumes. Voxelwise dynamic PET data analysis could instead provide a better insight into intra-tumour heterogeneity of gliomas. In this work, we investigate the ability of the widely used principal component analysis (PCA) method to extract meaningful quantitative dynamic features from high-dimensional motion-corrected dynamic [S-methyl-\textsuperscript{11}C]methionine PET data in a first cohort of 20 glioma patients. By means of realistic numerical simulations, we demonstrate the robustness of our methodology to noise. In a second cohort of 13 glioma patients, we compare the resulting parametric maps to these provided by standard one- and two-tissue compartment pharmacokinetic (PK) models. We show that our PCA model outperforms PK models in the identification of intra-tumour uptake dynamics heterogeneity while being much less computationally expensive. Such parametric maps could be valuable to assess tumour aggressiveness locally with applications in treatment planning as well as in the evaluation of tumour progression and response to treatment. This work also provides further encouraging results on the added value of dynamic over static analysis of [S-methyl-\textsuperscript{11}C]methionine PET data in gliomas, as previously demonstrated for O-(2-[\textsuperscript{18}F]fluoroethyl)-L-tyrosine.
\end{abstract}


\section{Introduction}
Gliomas are the most common primary brain tumours and are associated with poor prognosis. Glioma diagnosis and follow-up usually relies on magnetic resonance imaging (MRI), though addition of positron emission tomography (PET) with radio-labelled amino acids such as [S-methyl-\textsuperscript{11}C]methionine ([\textsuperscript{11}C]MET) has been shown to provide complementary information for tumour delineation \cite{lilja1985} and characterisation \cite{dewitte2001,sadeghi2007}, as well as for biopsy \cite{goldman1997,pirotte1997,pirotte2004} and therapy \cite{pirotte2009} planning. Whereas clinical amino-acid PET imaging of gliomas relies almost exclusively on static acquisitions, the added value of dynamic PET acquisitions has been demonstrated for tumour grading and genotyping \cite{poepperl2007,thon2014,jansen2015,roehrich2018,kunz2019,vettermann2019}, biopsy targeting \cite{thon2014}, and recurrence diagnosis \cite{pyka2018}. Aside from a longer acquisition time, the main limitation of dynamic PET imaging lies in the difficulty of extracting robust and clinically relevant features from noisy high-dimensional time-activity curves (TACs).

Previous works on dynamic PET imaging of gliomas with O-(2-[\textsuperscript{18}F]fluoroethyl)-L-tyrosine ([\textsuperscript{18}F]FET) -- another amino-acid PET tracer equivalent to [\textsuperscript{11}C]MET \cite{grosu2011} -- have highlighted differences in uptake dynamics between high-grade gliomas (HGGs) and low-grade gliomas (LGGs) by visually labelling mean tumour TAC as `increasing' or `decreasing.' It has been shown that a fast increasing then progressively decreasing mean TAC is characteristic of HGGs whereas a slowly increasing mean TAC is rather observed in LGGs \cite{poepperl2007,jansen2015}, and that foci with decreasing TAC should be taken into account for surgery guidance \cite{thon2014}. These interesting findings however have some limitations since TAC labelling does not allow continuous quantification of the dynamic behaviour and voxelwise extension might become challenging for large amounts of noisy data.

Recently, time-to-peak (TTP) has been investigated as a dynamic feature of interest for quantitative characterisation of mean TACs in glioma, with promising results for glioma grading \cite{jansen2015,roehrich2018,vomacka2018,kunz2019,vettermann2019} and recurrence diagnosis \cite{pyka2018} in dynamic [\textsuperscript{18}F]FET PET. TTP has the advantage of being easily computed and reflects to some extent the `increasing' or `decreasing' behaviour of TACs. However, this parameter is highly sensitive to data noise and depends on PET reconstruction framing, hence TTP values lie in a discrete range of arbitrarily chosen times.

Pharmacokinetic (PK) modelling is the gold standard method for dynamic PET data analysis. PK modelling relies on compartmental models whose kinetic parameters are estimated from the observed TACs given an arterial input function (AIF) (i.e the TAC of arterial blood used as an input for the model). This method has the great advantage of providing kinetic parameters directly linked to biological processes and has been previously used for glioma delineation in 2-deoxy-2-[\textsuperscript{18}F]fluoro-D-glucose \cite{spence2004} as well as for glioma genotyping in [\textsuperscript{18}F]FET PET \cite{roehrich2018, debus2018}. Nevertheless, direct compartment models fitting has two major limitations. First, it requires the user to provide an AIF, which can be either measured from arterial blood samples or extracted from large vessels appearing in the image. However, arterial sampling is an invasive procedure and is inconvenient in clinical practice. Image-derived input functions (IDIFs), on the other hand, are affected by partial volume effects inherent to PET imaging with direct impact on kinetic parameters estimation. Second, kinetic parameters fitting is a computationally expensive process and extension to voxelwise analysis may result in substantially long computation times for a single whole brain scan. Alternative methods have been proposed for PK analysis of dynamic PET data such as graphical analysis or reference tissue models, the latter not requiring an AIF. However, these methods generally provide only macro or relative kinetic parameters. A comparison of commonly used PK analysis methods for parametric map extraction from dynamic [\textsuperscript{18}F]FET PET data in diffuse gliomas has been recently performed by Koopman and colleagues \cite{koopman2020}.

Principal component analysis (PCA) is a commonly used unsupervised multivariate data analysis technique aiming at reducing high dimensional data space into a reduced number of components that best explain the observed data variance. Use of PCA for dynamic PET data analysis has been intensively studied by means of simulations, though surprisingly few clinical applications have seemed to emerge from these works. One major limitation of this technique is its limited ability to separate signal from noise for high noise levels or non-Gaussian noise distributions. Impact of noise level on PCA of dynamic PET images has been previously studied by Pedersen and colleagues \cite{pedersen1994}. They reported a strong influence of background noise on the computed components and highlighted the importance of per frame noise normalisation of the data. \v{S}\'amal and colleagues investigated the impact of data pre-processing on several PCA-related performance criteria using synthetic imaging datasets with different noise distributions \cite{samal1999}. They concluded that frame normalisation by noise variance was the optimum scaling for all noise types to separate signal from noise using PCA. However, non-normalised data were still found to provide satisfying results for both true and noisy signal reproduction.

To the best of our knowledge, most published clinical studies, including those mentioned above, are only concerned with the mean TAC inside one or several pre-delineated volume(s) of interest (VOI(s)). However, gliomas are known to be highly heterogeneous tumours that may exhibit multiple subregions with varying proliferation potential, aggressiveness, hypoxia level, and treatment-resistance abilities. Though computationally more expensive and more prone to noise, voxelwise analysis of dynamic PET data could instead provide a valuable intra-tumour insight. Voxelwise extensions of TTP analysis \cite{vomacka2018} and PK modelling \cite{debus2018} have been recently proposed but are still prone to the same limitations as their regionwise versions with increased influence of noise and stability issues reported for PK modelling \cite{debus2018}.

Furthermore, most dynamic PET studies rely on a limited number of non-uniformly sampled and non-overlapping frames with variable length. However, uniform TAC sampling with a large number of short frames could potentially benefit dynamic analysis by providing a higher temporal resolution and removing the need to select an arbitrary irregular framing, at the expense of data noise. In addition, high dimensional data with a large number of highly correlated features are particularly suitable for PCA. Moreover, reconstruction of overlapping frames leads to a good comprise between temporal resolution and count statistics and has been previously used for preclinical cardiac PET imaging \cite{territo2016}.

In this work, we investigate the ability of PCA to extract meaningful dynamic features from uniformly sampled high dimensional dynamic [\textsuperscript{11}C]MET PET data in 33 glioma patients. We assess the robustness of the proposed methodology to noise by means of realistic numerical simulations. We compare the derived parametric maps to these obtained from standard voxelwise PK modelling and conclude that PCA outperforms PK modelling in highlighting tumour sub-regions with different uptake dynamic behaviours. These results support the added value of dynamic over static analysis of [\textsuperscript{11}C]MET PET data in gliomas, as previously demonstrated for [\textsuperscript{18}F]FET.

\section{Methods}
\subsection{Image acquisition and reconstruction}
33 glioma patients (20 males, 28 surgically treated, with median age 55 yr) admitted to our institution for a diagnosis or follow-up [\textsuperscript{11}C]MET PET scan were enrolled in this study. Patient clinical data, 2016 WHO classification and undergone treatments at imaging time are provided for each lesion in \Cref{table_a1}. Patient cohort was further split into training ($n=20$, patients 1 to 20) and testing ($n=13$, patients 21 to 33) sets, respectively for PCA model building and evaluation (see below). All patients underwent a 30-minutes-and-30-seconds PET acquisition started 30 seconds before intravenous injection of 287 to 555 MBq of [\textsuperscript{11}C]MET. All acquisitions were performed on a Vereos digital PET-CT scanner (Philips Healthcare, The Netherlands) with an axial and trans-axial resolution of 4.1 mm at 1 cm from the field-of-view centre. 906 overlapping frames of 20 s, with a time spacing of 2 s and a voxel size of 2 mm$\,\times\,$2 mm$\,\times\,$2 mm were reconstructed from the LIST files using time-of-flight ordered-subset expectation maximisation (TOF-OSEM with 10 subsets and 3 iterations) and computed tomography (CT)-based attenuation correction (CTAC). No post-reconstruction filter was applied to the dynamic frames. A routine static PET image (20 to 27 min post-injection (p.i.)) was also reconstructed for each patient (2 mm$\,\times\,$2 mm$\,\times\,$2 mm -- CTAC TOF-OSEM with 15 subsets and 3 iterations -- no filter).

All experimental protocols in this study were approved by the Hospital-Faculty Ethics Committee of H\^opital Erasme (Accreditation 021/406), which waived the need for written consent (Ref. Q2017/020). All procedures performed in this study were in accordance with the 1964 Helsinki declaration and its later amendments or comparable ethical standards.

\subsection{Motion correction}
A systematic and progressive drift of the patients' head throughout the acquisition was observed in our dataset, imputed to the low stiffness of the scanner head support. Although motion between two consecutive frames is almost unnoticeable, it turned out that the patients' head had sunk of up to 1.5 mm over 30 minutes of acquisition. Such movement substantially impacts TACs at the voxel level and had to be corrected prior to analysis. Since frame-by-frame registration is not suitable for short frames with low signal-to-noise ratio (SNR), the following approach was used: Frames 36 to 906 (i.e. 40 s to 30 min p.i.) were grouped into consecutive blocks of 31 frames spaced by 30 frames. The mean frame of each block was computed and referred to as a `long frame.' The last long frame was used as a reference for rigid registration of the other long frames by mutual information maximisation \cite{maes1997}. Each computed rigid transform, stored as a versor, was assigned to the mid-frame time point of the corresponding long frame. Finally, for each short frame, a rigid transform at mid-frame time was linearly interpolated from the known long frame transforms, which can be trivially performed when expressing transforms as versors. All registrations were performed in Python using \textit{SimpleITK} \cite{lowekamp2013}.

\subsection{Image-derived input function extraction}
A blood input function was extracted from each registered volume. A vascular image (0 to 120 s p.i.) was first computed by averaging registered frames 16 to 66. Two regions of interest (ROIs) covering the petrous (horizontal) segment of each internal carotid artery was manually drawn on the early frame. The blood input function was obtained by averaging TACs of the 10 brightest voxels on the vascular image within both ROIs. Brighter voxels are indeed considered to be the less impacted by partial volume effect (PVE), as suggested previously \cite{parker2005,su2005}. Estimation of the spill-out effects introduced was performed both analytically and by means of numerical simulations for realistic internal carotid arteries dimensions and scanner resolution (see \Cref{section_a_2,section_b_1}). The estimated spill-out coefficient of 0.51 was used to correct the underestimated image-derived blood input functions.

\subsection{Time-activity curves extraction and registration}
TACs within the brain region (excluding skull and peripheral cerebrospinal fluid) were extracted from each registered dynamic volume. For each patient, a brain ROI was first defined as follows: A late frame (20 to 30 min p.i.) was computed by averaging registered frames 616 to 906. Patient's most recent T2 FLAIR MR image (no more than one month apart) was rigidly registered to the late frame by mutual information maximisation \cite{maes1997}. Brain volume was then segmented on the registered T2-FLAIR image using a combination of thresholdings and morphological operations, available as part of an in-house C++ software based on \textit{VTK} \cite{schroeder2010} and \textit{ITK} \cite{yoo2002}. TACs within the brain ROI were spatially smoothed for noise reduction by averaging within a 3 voxels$\,\times\,$3 voxels$\,\times\,$3 voxels neighbourhood. For inter-patient normalisation purpose, all smoothed TACs were converted to SUV units and temporally registered to account for inter-patient injection delay. Injection delays were computed by cross-correlation maximisation between each patient's IDIF and the most delayed IDIF observed, used as reference (patient 17). In total, 6,420,534 TACs were extracted from the 33 dynamic scans.

\subsection{Biological tumour volume delineation}
For each lesion, the biological tumour volume (BTV) was delineated on the average late frame (20 to 30 min p.i.) with a threshold of 1.6 times the normal brain uptake \cite{pauleit2005}, computed as the mean SUV within a spherical ROI with radius 1 cm placed in the contralateral hemisphere symmetrically to the lesion location with regard to the falx cerebri. Basal nuclei were manually removed when erroneously included in the BTV. All segmentations were reviewed and approved by an experienced nuclear medicine physician. 

\subsection{Pharmacokinetic modelling}
\subsubsection{1TC model}
The simplest compartmental model is the one-tissue compartment model (1TCM) which describes transport of the tracer from blood to tissues. 1TCM is illustrated in \Cref{figure_a1}. Transport processes involved are classically modelled by first-order kinetics, yielding:
\begin{equation}
\frac{dc_1(t)}{dt} = K_1 c_p(t) - k_2 c_1(t)
\label{equation_1}
\end{equation}
where $c_p(t)$ and $c_1(t)$ are respectively the tracer activity concentration of the plasma and tissue compartment at time $t$, and $K_1$ and $k_2$ are the transport rate constants from plasma to tissues and conversely. Calculating the unilateral Laplace transform of \Cref{equation_1} and considering zero initial concentration in both compartments leads to the expression of the transfer function $H_1(s)$ of the 1TCM:
\begin{equation}
sC_1(s) = K_1 C_p(s) - k_2 C_1(s) \quad \Leftrightarrow \quad H_1(s) = \frac{C_1(s)}{C_p(s)} = \frac{K_1}{s+k_2}
\label{equation_2}
\end{equation}
where $C_p(s)$ and $C_1(s)$ are the respective Laplace transforms of $c_p(t)$ and $c_1(t)$ for Laplace variable $s$. In this work, we propose the following change of variables:
\begin{equation}
G = \frac{K_1}{k_2}, \quad \tau = \frac{1}{k_2} \quad \Leftrightarrow \quad H_1(s) =\frac{G}{\tau s + 1}
\label{equation_3}
\end{equation}
This transformation is motivated by linear time-invariant system theory where $G$ and $\tau$ are commonly referred to as the system static gain and time constant. In a physiologic point a view $G$ is also referred to as the tracer total volume of distribution. $G$ thus reflects the steady-state tissue to plasma uptake ratio (i.e. the asymptotic uptake value for a unit step plasma input function) whereas $\tau$ describes the uptake dynamics off-steady-state (i.e. the shape of the tissue TAC) and should reflect the `increasing' or `decreasing' behaviour of tissue TACs. 

\begin{figure}[ht!]
\centering
\includegraphics[width=0.3\linewidth]{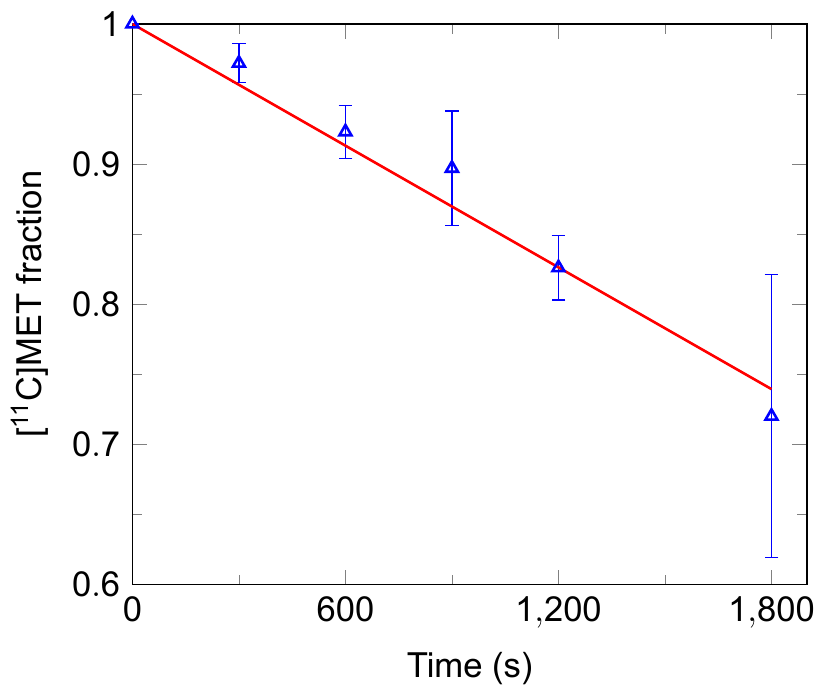}
\caption{Free plasma [\textsuperscript{11}C]MET fraction over time used for metabolite correction. Experimental population data represented by blue triangles with error bars (mean $\pm$ standard deviation) were obtained by Sato and colleagues from 18 glioma patients \cite{sato1992}. Least-squared fitted linear approximation is plotted in red.}
\label{figure_1}
\end{figure}

Since a certain fraction of blood vessels is comprised within a single PET voxel (around 5\% in healthy brain tissues), total voxel TAC is given by:
\begin{equation}
c_v(t) = \alpha c_b(t) + (1-\alpha) c_t(t)
\label{equation_4}
\end{equation}
where $\alpha$ is the vascular fraction and $c_v(t)$, $c_b(t)$ and $c_t(t)$ are respectively the voxel, blood and tissue total activity concentration, with $c_t(t) = c_1(t)$ for the 1TCM. Furthermore, $c_b$ can be expressed by:
\begin{equation}
c_b(t) = h c_e(t) + (1-h)c_{p_\text{tot}}(t)
\label{equation_5}
\end{equation}
where $h$ is the haematocrit (i.e. the erythrocyte volume fraction of blood), $c_e(t)$ is the erythrocyte activity concentration and $c_{p_\text{tot}}(t)$ is the total plasma activity concentration attributed to the tracer and its metabolites. Since no blood samples were available for the imaged patients and no population data were available for the erythrocyte uptake of [\textsuperscript{11}C]MET, $c_e(t)$ was neglected in first approximation. The haematocrit $h$ was set to its average population value of 0.45. The tracer plasma activity $c_p(t)$ was computed from the total plasma activity $c_{p_\text{tot}}(t)$ using the time-dependent parent fraction $f(t)$:
\begin{equation}
c_p(t) =  f(t) c_{p_\text{tot}}(t)
\label{equation_6}
\end{equation}
In this work, $f(t)$ was linearly interpolated from known population data points reported by Sato and colleagues \cite{sato1992} and depicted in \Cref{figure_1}. Taking into account the amount of time required for the tracer to flow from the carotid arteries -- where the blood input function is extracted -- to the voxel location, the voxel TAC is finally given by:
\begin{equation}
c_v(t) = \alpha c_b(t-d) + (1-\alpha) h_1(t) \ast \left(\frac{f(t-d)}{1-h}c_b(t-d)\right)
\label{equation_7}
\end{equation}
where $d$ is the system delay and $h_1$ is the inverse Laplace transform of $H_1$.

\subsubsection{2TC model}
The two-tissue compartment model (2TCM) introduces a second tissue compartment, representing either a physically separated compartment or a different state of the tracer (e.g. bound or metabolised). In the case of dynamic brain [\textsuperscript{11}C]MET PET imaging, tissue compartments 1 and 2 usually refer to extra and intra-cellular space respectively since dynamic changes in voxel activity concentration have been mainly attributed to tracer transport rather than metabolisation \cite{ishiwata1993}. Reversible 2TCM has been previously reported to best model [\textsuperscript{18}F]FET uptake in glioma \cite{koopman2018} and is illustrated in \Cref{figure_a1}. Transfer function for the 2TCM is derived in \Cref{equation_a16,equation_a17,equation_a18,equation_a19,equation_a20,equation_a21} (see \Cref{section_a_3}).

\subsubsection{Kinetic models fitting}
1TCM and 2TCM kinetic parameter values as well as vascular fraction $\alpha$ and delay $d$ were individually estimated for each voxel TAC in our dataset (33 patients, 6,420,534 curves) by least-squares transfer function fitting using the Python \textit{SciPy}'s `optimize' and `signal' modules \cite{virtanen2020}. The added value of overlapping frames over adjacent frames for 1TCM kinetic parameters fitting was investigated by means of numerical simulations (see \Cref{section_a_4,section_b_2}).

\subsection{Numerical simulations}
Numerical simulations were conducted to assess the impact of data noise on the ability of PCA to capture TAC dynamic behaviours. Synthetic time activity curves were generated using the 1TCM and the tri-exponential blood input function model proposed by Feng and colleagues \cite{feng1997}:
\begin{equation}
c_b(t) = (A_1t-A_2-A_3)e^{-\lambda_1t}+A_2e^{-\lambda_2t}+A_3e^{-\lambda_3t}
\label{equation_8}
\end{equation}
To be able to compute an analytical solution of the tissue tracer activity, a linear approximation of the time dependent parent fraction $f(t)=m\,t+p$ was built, leading to the following expression for the plasma input function:
\begin{equation}
c_p(t) = \frac{m\,t+p}{1-h} \left((A_1t-A_2-A_3)e^{-\lambda_1t}+A_2e^{-\lambda_2t}+A_3e^{-\lambda_3t}\right)
\label{equation_9}
\end{equation}
The value of $p$ was imposed such that free [\textsuperscript{11}C]MET plasma activity fraction is equal to 1 at injection time and the value of $m$ was least-squared fitted on the experimental data of Sato and colleagues\cite{sato1992}. The resulting linear approximation is depicted in \Cref{figure_1} (red line). Taking into account carotid-to-voxel delay $d$, the analytical expression of the 1TCM tissue tracer activity $c_t(t)$ for $t \geq d$ and the plasma input function in \Cref{equation_9} is given by:
\begin{align}
c_t(t) &= \frac{1}{1-h} \left[\frac{A_1 K_1}{k_2-\lambda_1} \left(m(t-d)^2 + \left(p+\frac{m}{k_2-\lambda_1}\right)(t-d)\right) e^{-\lambda_1 (t-d)} \right.
\nonumber \\
&+ \left. \sum_{i=1}^{3} \frac{\Tilde{A}_i K_1}{k_2-\lambda_i} \left[\left(p-\frac{m}{k_2-\lambda_i}\right)\left(e^{-\lambda_i (t-d)}-e^{-k_2 (t-d)}\right) + m(t-d)e^{-\lambda_i(t-d)}\right] \right.
\nonumber \\
&+ \left. \frac{A_1 K_1 m}{(k_2-\lambda_1)^3}\left(e^{-\lambda_1 (t-d)}-e^{-k_2 (t-d)}\right)\right]
\label{equation_10} \\
\Tilde{A} &= [-A_2-A_3-\frac{A_1}{K_1-\lambda_1}, A_2, A_3]
\nonumber
\end{align}
Mean voxel activity concentration $c_{v}^{t_s\rightarrow t_e}$ for a frame starting at time $t_s$ and ending at time $t_e$ is then given by:
\begin{equation}
c_{v}^{t_s\rightarrow t_e} = \frac{1}{t_e-t_s} \left(\alpha \int_{t_s}^{t_e} c_b(t-d) dt + (1-\alpha) \int_{t_s}^{t_e} c_t(t) dt\right)
\label{equation_11}
\end{equation}
where $c_b(t)$ and $c_t(t)$ are respectively given by \Cref{equation_8,equation_10} and:
\begin{align}
\int_{t_s}^{t_e} c_b(t-d) dt &= \frac{A_2+A_3-A_1\left(t_e-d+\frac{1}{\lambda_1}\right)}{\lambda_1} e^{-\lambda_1 (t_e-d)} - \frac{A_2+A_3-A_1\left(t_s-d+\frac{1}{\lambda_1}\right)}{\lambda_1} e^{-\lambda_1 (t_s-d)}
\nonumber \\
&- \sum_{i=2}^{3} \frac{A_i}{\lambda_i} \left(e^{-\lambda_i (t_e-d)} - e^{-\lambda_i (t_s-d)}\right)
\label{equation_12} \\
\int_{t_s}^{t_e} c_t(t) dt &= \frac{1}{1-h} \left[\frac{A_1 K_1}{k_2-\lambda_1}\left(\frac{m\lambda_1^2(t_s-d)^2 + \lambda_1\gamma(t_s-d)+\gamma}{\lambda_1^3}e^{-\lambda_1(t_s-d)} \right. \right.
\nonumber \\
&- \left. \left. \frac{m\lambda_1^2(t_e-d)^2 + \lambda_1\gamma(t_e-d)+\gamma}{\lambda_1^3}e^{-\lambda_1(t_e-d)}\right) \right.
\nonumber \\
&+ \left. \sum_{i=1}^{3} \frac{\Tilde{A}_i K_1}{k_2-\lambda_i} \left[\frac{p-\frac{m}{k_2-\lambda_i}}{k_2} \left(e^{-k_2(t_e-d)}-e^{-k_2(t_s-d)}\right) \right. \right.
\nonumber \\
&+ \left. \left. \frac{m\lambda_i(t_s-d) + \left(p-\frac{m}{k_2-\lambda_i}\right) \lambda_i + m}{\lambda_i^2}e^{-\lambda_i(t_s-d)} \right. \right.
\nonumber \\
&- \left. \left. \frac{m\lambda_i(t_e-d) + \left(p-\frac{m}{k_2-\lambda_i}\right) \lambda_i + m}{\lambda_i^2}e^{-\lambda_i(t_e-d)} \right] \right.
\nonumber \\
&+ \left. \frac{A_1 K_1 m}{(k_2-\lambda_1)^3} \left( \frac{e^{-k_2(t_e-d)}-e^{-k_2(t_s-d)}}{k_2} - \frac{e^{-\lambda_1(t_e-d)}-e^{-\lambda_1(t_s-d)}}{\lambda_1} \right) \right]
\nonumber \\
\label{equation_13} \\
\gamma &= \left(p+\frac{m}{k_2-\lambda_1}\right)\lambda_1 + 2m
\nonumber
\end{align}
For each simulated frame, synthetic noise was generated using the model proposed by Logan and colleagues \cite{logan2001}:
\begin{equation}
c_{v,\text{noisy}}^{t_s\rightarrow t_e} = c_{v}^{t_s\rightarrow t_e} + \beta\sqrt{\frac{c_{v}^{t_s\rightarrow t_e}}{(t_e-t_s)e^{-\lambda \frac{t_e-t_s}{2}}}}G(0,1)
\label{equation_14}
\end{equation}
where $c_{v,\text{noisy}}^{t_s\rightarrow t_e}$ is the noisy mean voxel tracer activity concentration for a frame starting at time $t_s$ and ending at time $t_e$, $\beta$ is a scaling factor, $\lambda$ is the isotope decay constant and $G(0,1)$ is a random number generated from a Gaussian distribution with mean 0 and standard deviation 1.

Synthetic TACs were generated using \Cref{equation_11,equation_12,equation_13} from the 1TCM kinetic parameter values previously fitted on each real TAC in our dataset (33 patients, 6,420,534 TACs). The input function parameter values used in \Cref{equation_8} were least-squared fitted for each corresponding patient's IDIF using the Python \textit{SciPy}'s `optimize' and `signal' modules \cite{virtanen2020}. The generated synthetic dataset is thus similar to the real dataset but has known ground truth signal. Synthetic noise was added to simulated TACs using \Cref{equation_14} for increasing noise levels $\beta$ ranging from 0.25 to 2.0 with step 0.25.

\subsection{Principal component analysis}
Principal component analysis is a commonly used dimension reduction technique that aims at finding the linear transformation of the initial data space into the so-called `components' space which best explains the data variance. More formally, let $\boldsymbol{X}$ be the data matrix of dimensions $n\times p$ where $n$ is the number of observations (i.e. the number of TACs) and $p$ the number of variables (i.e. the number of samples per TAC). The method aims at finding the new axis system $\left(\boldsymbol{e_1}, \boldsymbol{e_2}, \dots, \boldsymbol{e_p}\right)$ in which the variance
\begin{equation}
s^2_{\boldsymbol{e_i}} = \boldsymbol{e_i}^\top \boldsymbol{S} \boldsymbol{e_i}
\label{equation_15}
\end{equation}
is maximised under constraints $\boldsymbol{e_i}^\top \boldsymbol{e_j} = \delta_{i,j}$ and $s^2_{\boldsymbol{e_1}} \geq s^2_{\boldsymbol{e_2}} \geq \dots \geq s^2_{\boldsymbol{e_p}}$, where $\delta_{i,j}$ is the Kronecker delta and $\boldsymbol{S}$ is the data covariance matrix given by:
\begin{equation}
\boldsymbol{S} = \frac{1}{n-1}\boldsymbol{X_c}^\top \boldsymbol{X_c}
\label{equation_16}
\end{equation}
where $\boldsymbol{X_c}$ is the centred data matrix, i.e. the data matrix from which variable means have been subtracted columnwise. It can be shown that such components are given by the eigen vectors of $\boldsymbol{S}$ ordered by decreasing eigenvalues, corresponding to the respective component variances. As the amount of explained data variance decreases with the component number, it is expected that a sufficiently high amount of data variance can be explained by a reduced number of components $m \ll p$, which is the case for highly correlated initial variables.

As previously stated, our real and synthetic TAC datasets were split into training and testing sets on a patient basis to avoid over-fitting bias in the evaluation of PCA performance. The training and testing datasets comprise TACs from 20 (patients 1 to 20) and 13 (patients 21 to 33) patients respectively, for a total of 3,974,466 and 2,446,068 curves.

Impact of noise on true signal reconstruction as well as on the first six principal components was first investigated using the synthetic TAC dataset. For each level of noise, principal components were determined on the corresponding noisy training synthetic TAC set (20 patients, 3,974,466 TACs) and true signal reconstruction with 4 and 6 components was assessed on the test synthetic TAC set (13 patients, 2,446,068 TACs). To this end, component values were evaluated for each noisy test synthetic TAC at the considered noise level and estimated denoised TACs were reconstructed from the first 4 and 6 component values, then compared to the corresponding true unnoisy synthetic TAC using mean squared error. Influence of noise level on the components themselves was also assessed by computing their cosine with the corresponding component obtained for a noise level of zero. 
The first 6 principal components were then determined on the real training TAC set (20 patients, 3,974,466 TACs) and compared to these determined on the synthetic training dataset for a noise level of zero by computing of their respective cosine. Component values computed over the TACs of the 13 test patients were mapped spatially to their respective voxel locations and the resulting PC maps were compared to the corresponding 1TCM and 2TCM kinetic parameter maps, visually and by means of Spearman's correlation analyses for the 1TCM parametric maps.

Due to the limited amount of memory available, the incremental PCA (IPCA) algorithm was used for all analyses instead of matrix decomposition since it allows a batch processing of the data. The implementation used is available in Python as part of the \textit{scikit-learn}'s `decomposition' module \cite{pedregosa2011}. No noise normalisation was performed prior to PCA.

\section{Results}
\subsection{Principal components}
\subsubsection{Synthetic data}
\begin{figure}[ht!]
\centering
\includegraphics[width=0.9\linewidth]{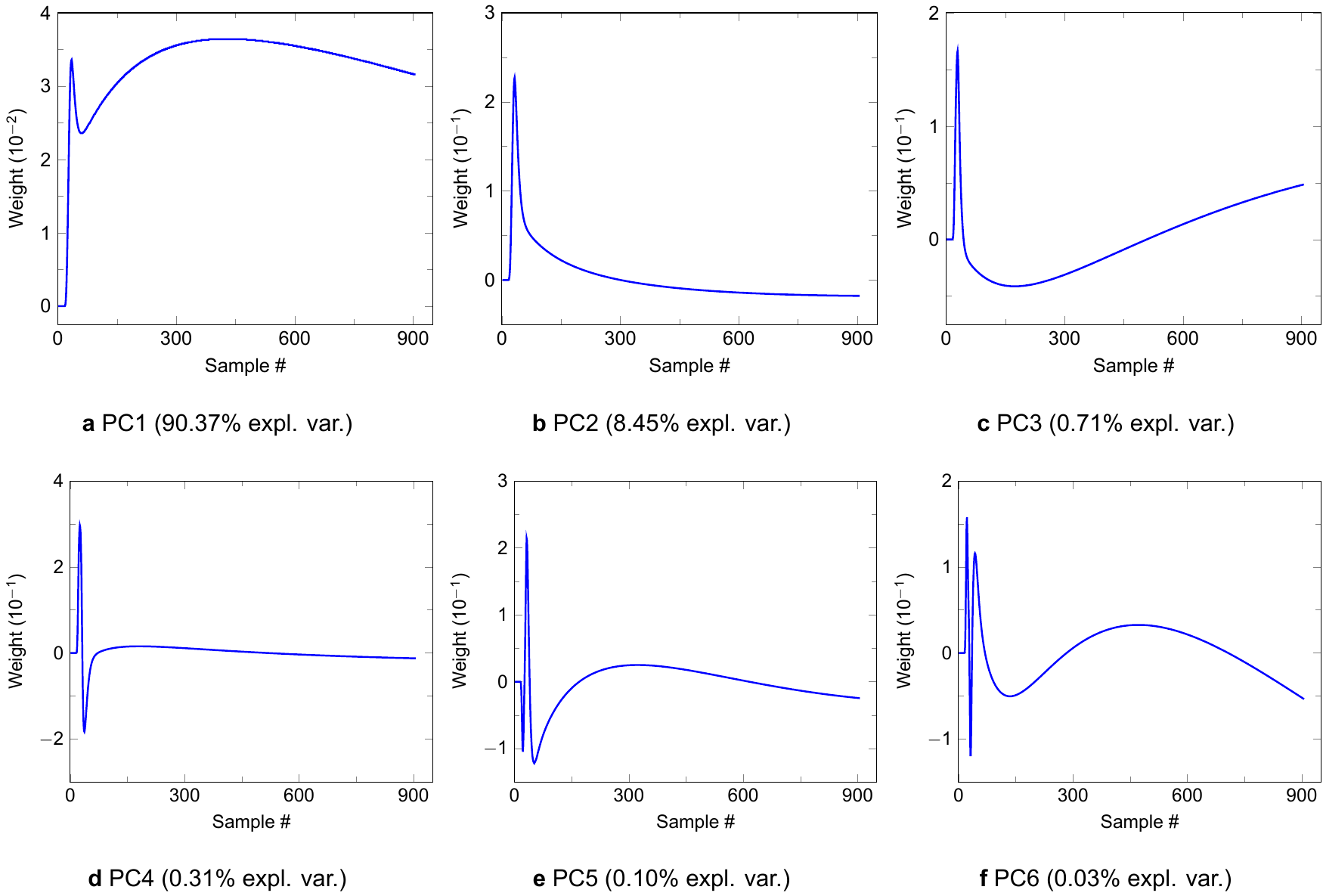}
\caption{First six principal components (PCs) with their explained variance ratio determined on the training synthetic dataset (20 patients, 3,974,466 TACs) for a noise level of zero. Each component assigns a weight to each of the 906 TAC samples, starting from 10 s to 1820 s with step 2 s.}
\label{figure_2}
\end{figure}

The first six principal components (PCs) determined on the training synthetic dataset (20 patients, 3,974,466 TACs) with a noise level of zero are depicted in \Cref{figure_2}. Their corresponding explained variance ratios are respectively 90.37\%, 8.45\%, 0.71\%, 0.31\%, 0.10\%, and 0.03\% for PC1 to 6, totalling 99.97\% of the explained variance. It should however be noted that PC5 and PC6 have very low contribution to the explained variance. Clinical interpretation of the PCs can be made based on to their shape. Aside from the early peak, PC1 (\Cref{figure_2}\textbf{a}) assigns a relatively constant weight to the late samples and thus partly reflects mean tracer uptake. PC2 (\Cref{figure_2}\textbf{b}) overweights the early samples, then assigns a rapidly decaying weight to the next samples with a slightly negative value for the last samples. This component thus has a high value for TACs with high early activity as observed in blood after bolus injection of the tracer. PC3 (\Cref{figure_2}\textbf{c}) assigns a negative weight to the first half of the samples and a positive weight to the second half, with a quasi-linear increase. It thus has a negative value for decreasing TACs, a positive value for increasing TACs and a value around zero for flat TACs. PC4 (\Cref{figure_2}\textbf{d}) assigns a strongly positive weight to the very first samples, a strongly negative weight to the next few samples then a small quasi-constant weight to the last samples. PC4 thus assigns a negative value to TACs with delayed arterial peak and a positive value to TACs with non-delayed arterial peak. PC5 and 6 (\Cref{figure_2}\textbf{e} and \textbf{f}) are less easily interpreted but together only explain 0.13\% of the variance.

\begin{figure}[ht!]
\centering
\includegraphics[width=0.6\linewidth]{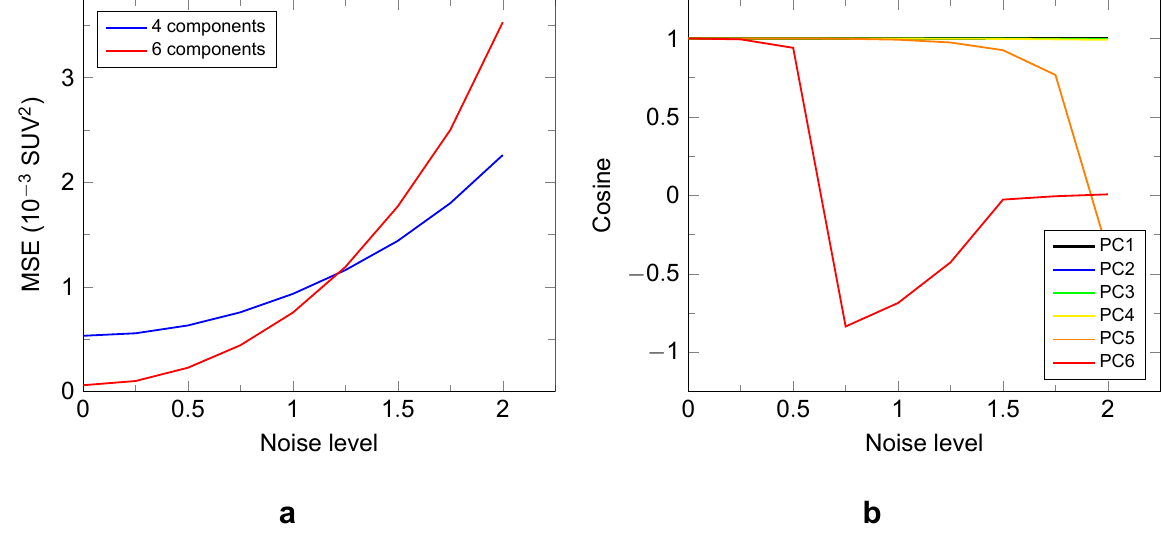}
\caption{Influence of the noise level on true signal reconstruction and on the principal components for the synthetic dataset. \textbf{a}: Mean squared error (MSE) computed between the true and estimated unnoisy test TACs reconstructed with 4 (blue curve) and 6 (red curves) components for different noise levels. \textbf{b}: Cosines between each principal component computed for noise levels ranging from 0.25 to 2.0 with step 0.25 and the corresponding component computed for a noise level of zero.}
\label{figure_3}
\end{figure}

\begin{figure}[ht!]
\centering
\includegraphics[width=0.9\linewidth]{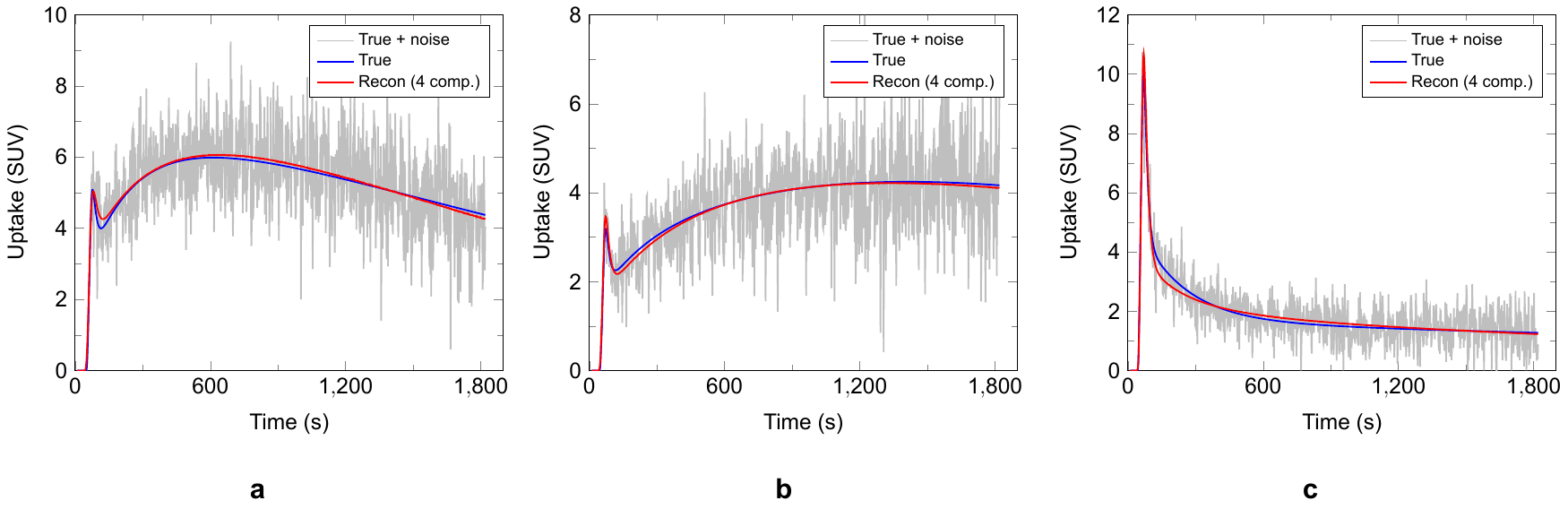}
\caption{Examples of PCA reconstructed synthetic test TACs with 4 components for a realistic noise level of 1.5. Noisy synthetic TAC is depicted in light grey along with the true unnoisy TAC (blue) and PCA denoised TAC with 4 components (red).}
\label{figure_4}
\end{figure}

\Cref{figure_3}\textbf{a} depicts the influence of the noise level on the unnoisy signal reconstruction assessed by mean squared error on the synthetic test TAC set (13 patients, 2,446,068 TACs). Interestingly, signal reconstruction with 6 components is less accurate than that with 4 components for noise levels above 1.25. \Cref{figure_3}\textbf{b} depicts the impact of noise on the first six computed principal components, assessed by their cosine with the respective first six principal components computed for a noise level of zero, used as reference and depicted in \Cref{figure_2}. PC5 cosine progressively drops from a noise level of 1.0 whereas a rapid drop of PC6 is observed for noise levels above 0.5. PC1 to 4, on the other hand, remain quite stable even for high noise levels. Examples of true unnoisy, noisy and PCA denoised test TACs with 4 components are depicted in \Cref{figure_4}. Reconstruction with only 4 components is remarkably accurate while efficiently removing noise.

\subsubsection{Real data}
\begin{figure}[ht!]
\centering
\includegraphics[width=0.9\linewidth]{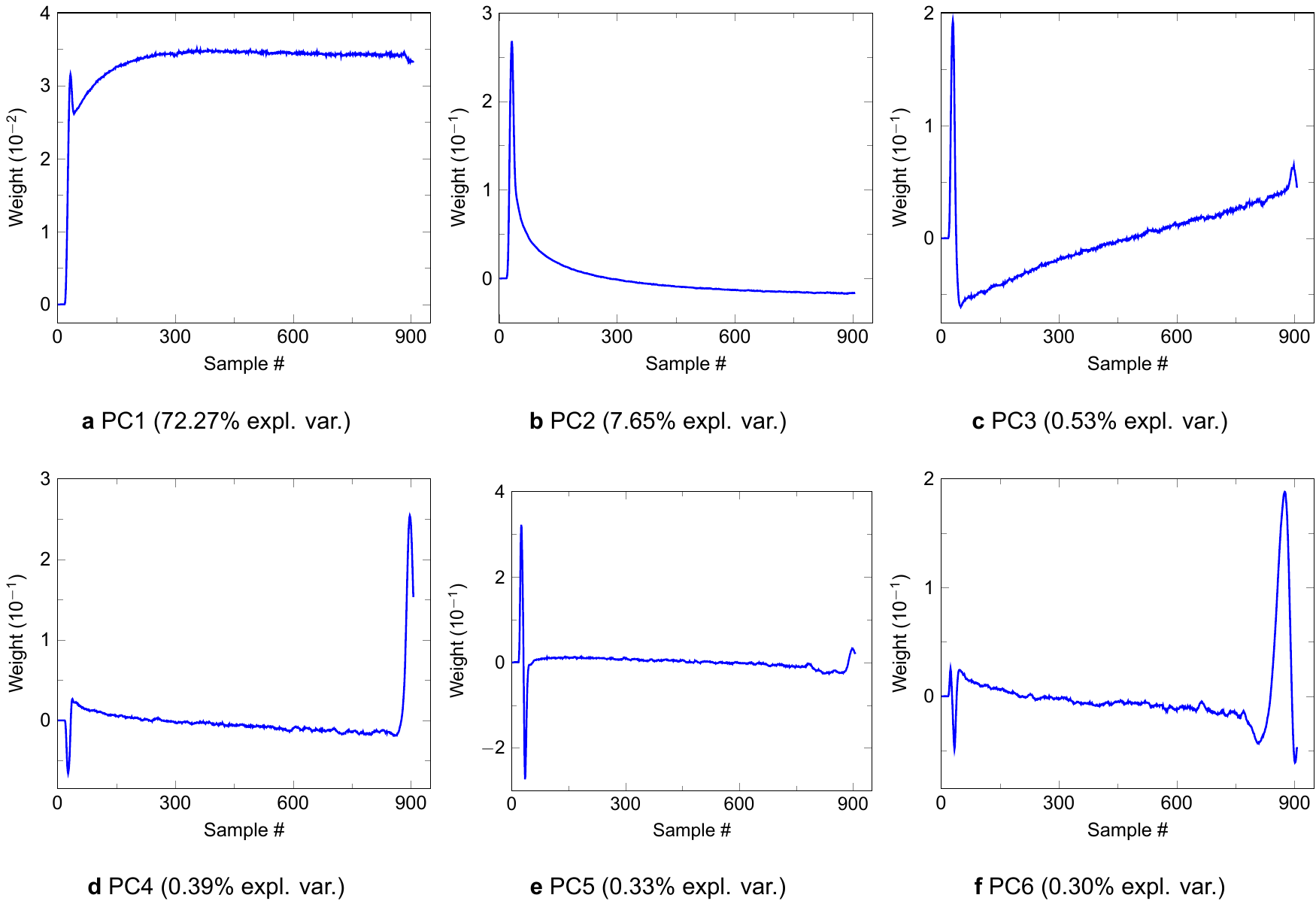}
\caption{First six principal components (PCs) with their explained variance ratio determined on the training real dataset (20 patients, 3,974,466 TACs). Each component assigns a weight to each of the 906 TAC samples, starting from 10 s to 1820 s with step 2 s.}
\label{figure_5}
\end{figure}

The first six principal components determined on the real training dataset are depicted in \Cref{figure_5}. Their corresponding explained variance ratios are respectively 72.27\%, 7.65\%, 0.53\%, 0.39\%, 0.33\%, and 0.30\% for PC1 to 6, totalling  81.47\% of the explained variance. Strong similarities are respectively observed between PC1-4 from the synthetic dataset (see \Cref{figure_2}) and PC1-3 and 5 from the real dataset (see \Cref{figure_5}), hence the same clinical interpretations can be made for the latter. These similarities are confirmed by their respective cosines of 0.9987, 0.9891, 0.9383 and 0.9111. The residual differences are suspected to originate from (i) the limited ability of the 1TCM used for data synthesis to fully capture the whole range of observed TAC dynamic behaviours and (ii) unmodelled additional noise sources and artefacts related to data reconstruction and residual patient motion. It should be noted that PC4-6 from the real dataset have very similar explained variance ratios, hence their order is not informative and could have been reversed for another similar dataset. Since PC4 and 6 from the real dataset are hardly interpretable and does not match any of the first six synthetic PCs in contrast to PC5, they will not be further considered in the following.

\subsection{Parametric maps}
\begin{figure}[ht!]
\centering
\includegraphics[width=\linewidth]{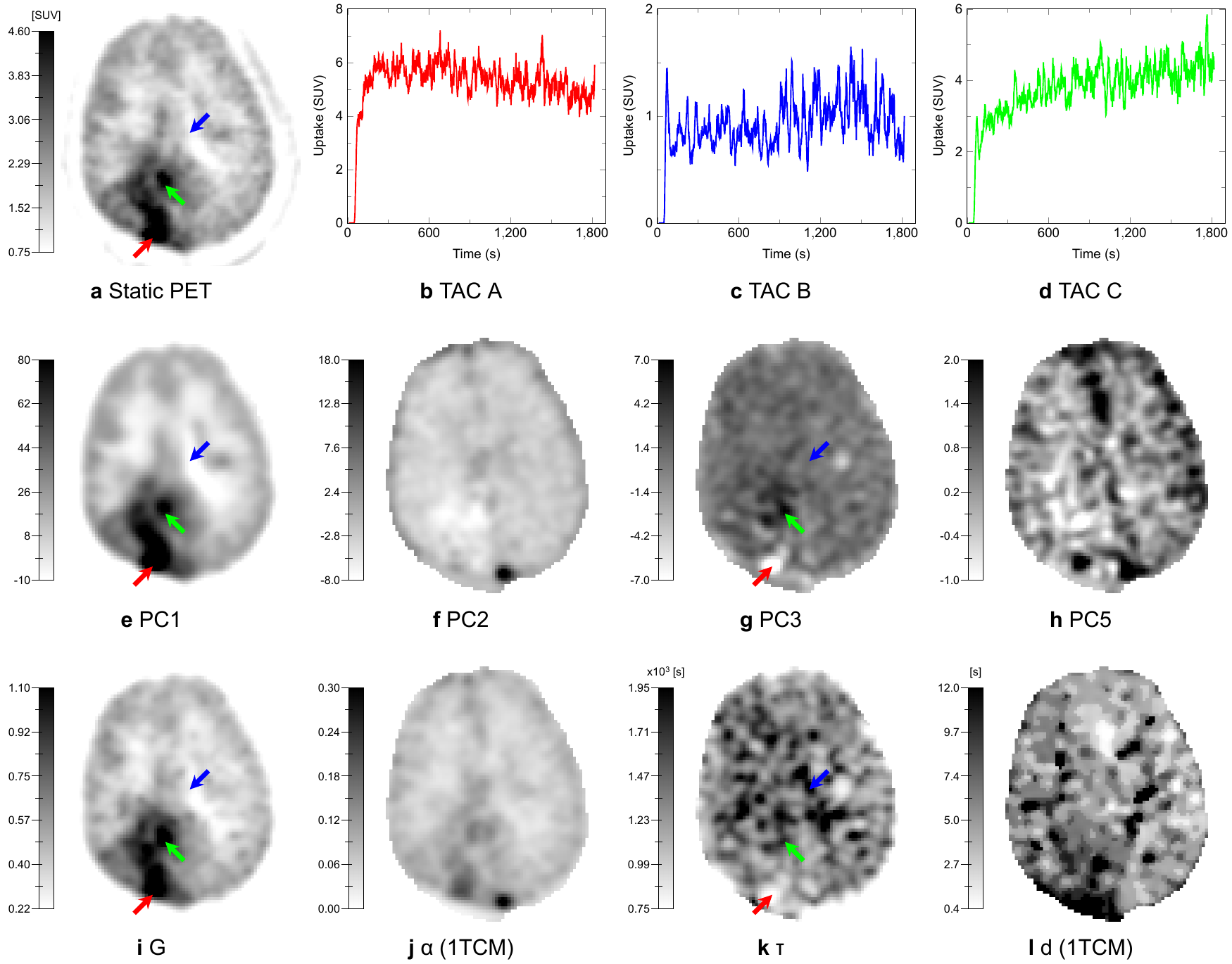}
\caption{Parametric maps generated from a dynamic [\textsuperscript{11}C]MET PET scan of a glioblastoma test patient (patient 21). \textbf{a}: Static PET (20 to 27 min p.i.). \textbf{b}-\textbf{d}: Smoothed TACs at voxels pointed by the red, blue and green arrows respectively. TAC A, B and C respectively have a `decreasing', `flat', and `increasing' behaviour. \textbf{e}-\textbf{h}: Mapped values of the principal components 1-3 and 5 from the real dataset. \textbf{i}-\textbf{l}: 1TCM kinetic parametric maps. All maps are displayed in inverted greyscale.}
\label{figure_6}
\end{figure}

In the following, PC definitions used are these determined on the real training dataset (see \Cref{figure_5}). Parametric maps obtained by voxelwise mapping of the PC values and 1TCM kinetic parameters computed from the voxel TACs of a glioblastoma test patient (patient 21) are depicted in \Cref{figure_6} in inverted greyscale. Static PET image (20 to 27 min p.i.) is depicted in \Cref{figure_6}\textbf{a}. Curves A, B and C (\Cref{figure_6}\textbf{b} to \textbf{d}) correspond to the smoothed TACs at voxels pointed by the red, blue, and green arrows respectively. PC1 map \Cref{figure_6}\textbf{e}) exhibits strong similarities with static PET (\Cref{figure_6}\textbf{a}) and $G$ map (\Cref{figure_6}\textbf{i}). PC2 map (\Cref{figure_6}\textbf{f}) exhibits strong similarities with the $\alpha$ (1TCM) map (\Cref{figure_6}\textbf{j}). PC3 map (\Cref{figure_6}\textbf{g}) and $\tau$ map (\Cref{figure_6}\textbf{k}) share visual similarities but structures are hardly distinguishable on the $\tau$ map as opposed to PC3. PC5 (\Cref{figure_6}\textbf{h}) and $d$ (1TCM) maps (\Cref{figure_6}\textbf{l}) exhibit fairly similar patterns with inverted contrast for PC5. 2TCM parametric maps are available for the same patient in \Cref{figure_a4} (see also \Cref{section_b_3}).

Furthermore, PC3 map clearly highlights tissue regions with different kinetic behaviours, as illustrated by TACs A, B, and C. Voxels with fast increasing then decreasing TACs appear brighter in PC3 map (red arrow), whereas progressively increasing TACs appear darker in PC3 map (green arrow). Voxels with relatively flat TACs appear in medium grey value (blue arrow). In contrast, the $\tau$ map fails to clearly highlight these different kinetic behaviours, as illustrated by voxels pointed by the green and blue arrows, both appearing darker in \Cref{figure_6}\textbf{k}. 

\begin{figure}[ht!]
\centering
\includegraphics[width=\linewidth]{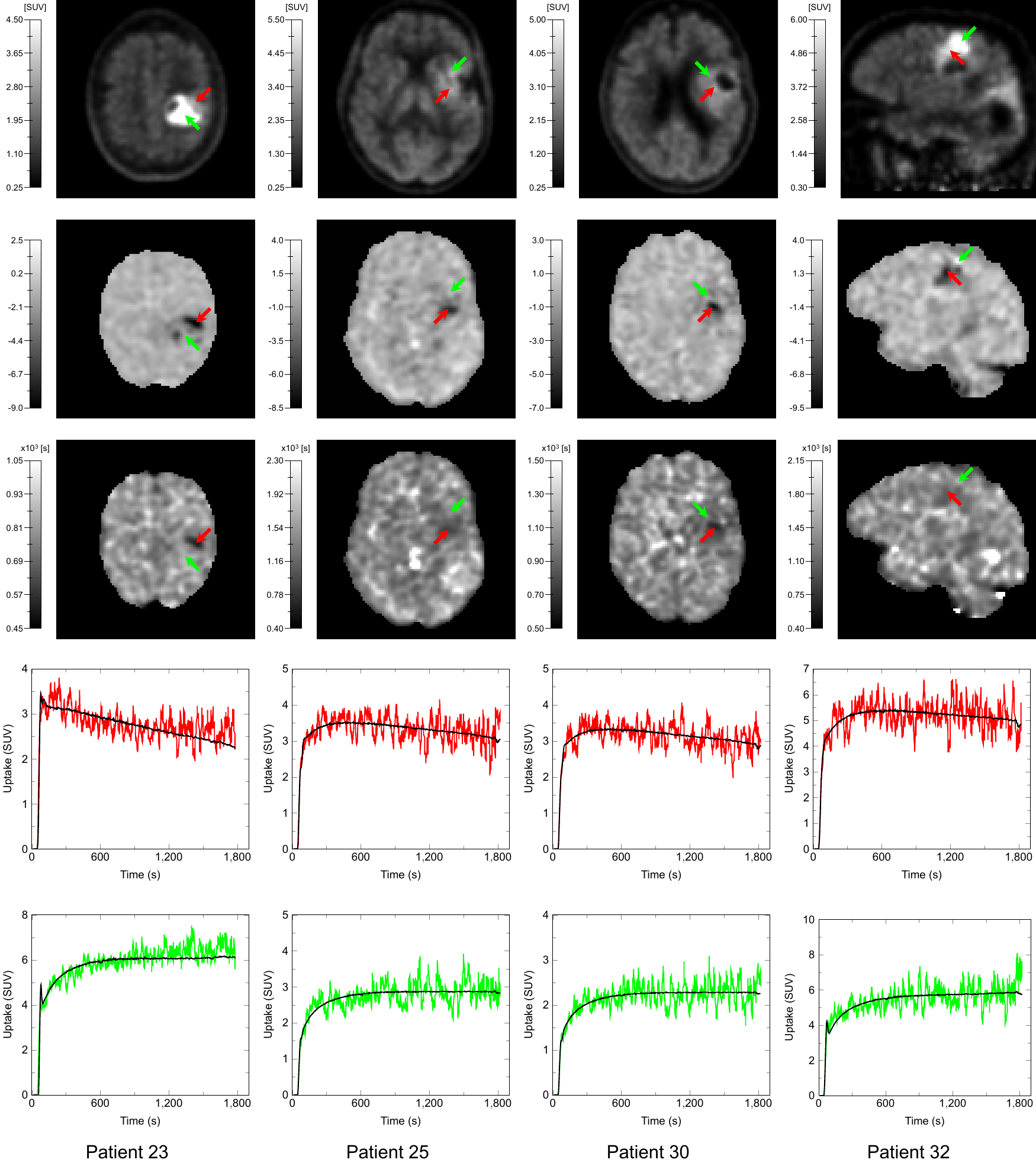}
\caption{Static PET image (20 to 27 min p.i. -- 1\textsuperscript{st} row), PC3 map (2\textsuperscript{nd} row) and $\tau$ map (3\textsuperscript{rd} row) derived from dynamic [\textsuperscript{11}C]MET scans of four test patients (patients 23, 25, 30 and 32) along with smoothed TACs at voxels pointed by the red (4\textsuperscript{th} row) and green (5\textsuperscript{th} row) arrows respectively. Red TACs have a `decreasing' or `flat' behaviour whereas green TACs have a `flat' or `increasing' behaviour. TAC reconstruction with four principal components (PC1-3 and 5) is superimposed in black to each raw TAC.}
\label{figure_7}
\end{figure}

Static PET images (20 to 27 min p.i.), PC3 and $\tau$ maps along with typical smoothed TACs at voxels pointed by the red and green arrows are depicted in \Cref{figure_7} for 4 additional test patients (patients 23, 25, 30, and 32 -- see \Cref{table_a1}). As opposed to static PET image, PC3 map allows to distinguish between voxels with `decreasing' or `flat' TACs (red) and voxels with `flat' or `increasing' TACs (green). $\tau$ maps exhibits similar patterns as PC3 maps within the BTV but with a substantially lower contrast. Noteworthy, similar behaviours were observed for the other 8 test patients. Interestingly, for the green TAC of patient 23 in \Cref{figure_7}, late uptake increase is not totally captured by PC1-3 and 5 (black curve) but PC4 and 6 are required for a more accurate reconstruction. 

Biological tumour volume (BTV), mean and maximum tumour-to-background ratio (TBR) evaluated on the static PET image (20 to 27 min p.i.) as well as tumour contrast (see \Cref{equation_a22}) evaluated on the static PET image and on the PC1 map are reported for each lesion in \Cref{table_a3}. Minimum, maximum and mean values of the 1TCM kinetic parameters and PCs within the BTV are respectively reported for each lesion in \Cref{table_a4,table_a5}.

\subsection{Features correlation}
Similarities between PCs and 1TCM kinetic parameter maps are confirmed by the pairwise Spearman’s correlation coefficients computed individually for each of the 13 test patients and summarised in \Cref{table_1}. Feature pairs PC2/$\alpha$ and PC5/$d$ strongly correlate whereas PC1/$G$ and PC3/$\tau$ very strongly correlate. Moderate positive correlations are also found between PC1/$\alpha$ -- imputed to the overweighting of the very first samples of PC1 (see \Cref{figure_5}\textbf{a}) -- and PC3/$G$.

\begin{table}[ht!]
\centering
\begin{tabular}{l|rrrr|}
\cline{2-5}
 & \multicolumn{1}{c}{\textbf{$G$}} & \multicolumn{1}{c}{\textbf{$\alpha$}} & \multicolumn{1}{c}{\textbf{$\tau$}} & \multicolumn{1}{c|}{\textbf{$d$}} \\ \hline
\multicolumn{1}{|l|}{PC1} & $0.90 \pm 0.04$ & $0.56 \pm 0.06$ & $0.05 \pm 0.06$ & $-0.02 \pm 0.05$ \\
\multicolumn{1}{|l|}{PC2} & $-0.30 \pm 0.19$ & $0.70 \pm 0.14$ & $-0.34 \pm 0.05$ & $-0.17 \pm 0.11$ \\
\multicolumn{1}{|l|}{PC3} & $0.49 \pm 0.13$ & $0.11 \pm 0.14$ & $0.86 \pm 0.05$ & $-0.26 \pm 0.11$ \\
\multicolumn{1}{|l|}{PC5} & $-0.07 \pm 0.08$ & $0.26 \pm 0.20$ & $-0.31 \pm 0.08$ & $-0.69 \pm 0.09$ \\ \hline
\end{tabular}
\caption{Spearman's correlation coefficients (median $\pm$ median absolute deviation) of all pairs of principal component / kinetic parameter maps computed individually for each of the 13 test patients.}
\label{table_1}
\end{table}

\section{Discussion}
We showed the ability of PCA to accurately capture different dynamic behaviours from a broad spectrum of high dimensional [\textsuperscript{11}C]MET PET TACs extracted from the whole brain region of 20 patients, most of whom have been surgically treated. By means of realistic numerical simulations, we demonstrated the robustness to noise of the first four principal components and validated the accuracy of the unnoisy signal reconstruction from the 1TCM kinetic parameter values of 13 other patients, without the need for prior noise normalisation. A possible explanation for this robustness could be that only TACs within brain region were considered for this work, hence background noise had no influence the computed components, as opposed to the works of Pedersen and colleagues \cite{pedersen1994} and \v{S}\'amal and colleagues \cite{samal1999}.

Four components (PC1-3 and 5) among the six that best explain variations observed in real whole brain TACs of 20 patients respectively matched the first four principal components derived from our synthetic training dataset. These four components have been found to be of clinical interest. PC1 provides a contrast similar to the routinely acquired static PET image (20 to 27 min p.i.) but has two advantages over the latter: (i) PC1 map has a higher tumour contrast (see \Cref{equation_a22}) than the static PET image, as reported in \Cref{table_a3} and (ii) PC1 does not depend on an arbitrary chosen imaging timing since each of the acquired samples from tracer injection contributes to some extent to the component value, making it more robust to inter-protocol variations. PC2 reflects the voxel vascular fraction but its small negative weighting of the late samples (see \Cref{figure_5}\textbf{b}) makes interpretation less straightforward in high uptake tumour regions. PC3 provides an interesting dynamic contrast related to the shape of the tissue TAC, distinguishing voxel with fast increasing then decreasing, flat or progressively increasing TACs. PC5 seems related to the carotid-to-voxel blood delay and could reflect the effectiveness of micro-vascularisation, which is known to be impaired in gliomas \cite{hardee2012}.

These four PCs strongly to very strongly correlate with the kinetic parameters of the widely used 1TCM PK model, which partly confirms our intuition concerning their clinical interpretation. On the other hand, correlation between kinetic parameters and PCs highlights the ability of a simplistic model such as the 1TCM to accurately capture the observed TACs variability. PCA however has two advantages over the 1TCM. First, it does not require a blood input function, avoiding invasive arterial sampling procedures or manual extraction of an IDIF potentially affected by PVE. Second, PCA parametric maps are much less computationally expensive to compute, with processing times of a few seconds for any new dynamic volume versus more than 3 hours of fit for the 1TCM on a high-end computer. Moreover, PC3 map has been shown to outperform the 1TCM $\tau$ map in differentiating voxels with `increasing' and `decreasing' TAC. In turn, the 2TCM did not provide any further meaningful parametric map for assessing heterogeneous dynamic behaviour within tumours (see \Cref{section_b_3}). Instability of 2TCM parameters fitting has been previously reported by Debus and colleagues \cite{debus2018} and could result from insufficient scan time. This combined with even longer processing times compared to the 1TCM leads us to conclude that the 2TCM is not suitable for voxelwise analysis of dynamic [\textsuperscript{11}C]MET PET data. Alternatively, a basis function approach for 1TCM and 2TCM parameters fitting could be used to shorten processing and improve stability \cite{gunn1997} but at the expense of precision on the kinetic parameter values and still longer processing times than PCA.

However, PCs must not been seen as a surrogate for the kinetic parameters since each PC may reflect a mixture of the kinetic constants and other parameters. PCs instead capture the largest variations observed among TACs resulting from complex underlying biological processes, under orthogonality constraint. Unequivocal link between one PC and one biologically relevant micro-parameter is not guaranteed either, nor is it the case for kinetic parameters, which both capture complex amino acid transport processes influenced, among others, by the number of amino-acid transporters on the endothelial and tumour cells membrane, the concentrations of all endogenous transporter shared substrates in every model compartment and the relative sizes of the compartments and thus (over-)cellularity \cite{panitchob2016}.

These findings may nevertheless have important implications for clinical management of gliomas: PCA applied to dynamic [\textsuperscript{11}C]MET PET data is likely to provide additional quantitative spatial information on glioma heterogeneity with little modification of the routine acquisition protocol, that is a longer acquisition time on the PET/CT tomograph but identical total duration of the procedure for the patient. In particular, PC3 map provides a novel contrast complementary to static PET image that can help distinguishing voxels with similar late uptake values but different uptake time courses. This component translates the `increasing' and `decreasing' behaviour of TACs observed of P\"opperl and colleagues at the whole tumour level \cite{poepperl2007} into a quantitative voxelwise metric. This metric might partly reflect biologically relevant parameters such as over-expression of LAT1, an important amino acid transporter over-expressed in gliomas at both blood-brain barrier and tumour cell level \cite{okubo2010, haining2012} transporting both [\textsuperscript{11}C]MET \cite{okubo2010, moulin-romsee2007} and [\textsuperscript{18}F]FET \cite{moulin-romsee2007}. LAT1 being an obligatory exchanger, its over-expression may indeed be associated with an increase in both influx and efflux of the tracer \cite{moulin-romsee2007}, leading to faster uptake dynamics that more closely follows the arterial input signal, hence characterised by a lower PC3 value. If this hypothesis turned out to be verified, application of the proposed methodology to dynamic [\textsuperscript{18}F]FET data would also be of great interest with an even more pronounced expected effect since this tracer is more specifically transported by LAT1 \cite{habermeier2015} and is not incorporated into proteins \cite{grosu2011,moulin-romsee2007}. However, the added value of this new contrast must still be validated by means of targeted biopsies. Relations between patient outcome or glioma molecular features such as IDH or 1p/19q codeletion status and PC values distribution within the BTV should also be investigated as part of a large scale study but was out of the scope of this preliminary methodological work. To the best of our knowledge, this is the first work to promote the added value of dynamic PET acquisitions with [\textsuperscript{11}C]MET in glioma patients to assess intra-tumour heterogeneity, as previously demonstrated for [\textsuperscript{18}F]FET.

Pharmacokinetic analyses conducted as part of this work were however prone to several limitations. Indeed, decision was made for this study not to perform arterial sampling which requires a dedicated device and poses risks for the patient and the nursing staff. Instead, blood input functions used for kinetic analyses were extracted from the image within the petrous segment of the internal carotid arteries. As a consequence, extracted input functions were affected by PVE and tracer metabolite activity, which could not be evaluated individually for each patient. Population-based haematocrit and metabolite correction and model-based spill-out coefficient estimation were proposed to counteract these effects but these corrections are still limited. Moreover, spill-in correction could not be performed as this is still a challenging open problem in PET imaging \cite{arkele2020}, which was out of the scope of this work. Nevertheless, it turned out in the course of this study that the introduction of the proposed input function corrections only affected the absolute values of the fitted kinetic parameters but that the contrast of the kinetic parametric maps was preserved. Since we were interested in this study in relative quantification between voxels to highlight intra-tumour heterogeneity in uptake dynamics, residual uncertainties in the corrected image-derived input functions were not suspected to significantly impact the conclusions of this work regarding the superiority of PCA to this extent.

Other dimension reduction techniques have been proposed in the literature for the analysis of dynamic PET data, such as factor analysis of dynamic structures (FADS) and non-negative matrix factorisation (NMF) \cite{cavalcanti2019}. These are distinguished by their ability to isolate signal from noise, which strongly depends on the assumptions made on noise distribution in PET images. Whereas noise related to event counting in PET imaging follows a Poisson distribution, noise in reconstructed PET images is less well characterised due to alterations related to system hardware and reconstruction algorithm, including scatter and attenuation corrections \cite{cavalcanti2019}, hence remains an open problem. Comparison of these methods in the particular case of dynamic [\textsuperscript{11}C]MET PET data of glioma patients would be of interest but is out of the scope of this paper. 

In conclusion, we showed the ability of PCA to extract meaningful parametric maps from noisy high dimensional dynamic [\textsuperscript{11}C]MET PET scans of glioma patients. One of these maps was found to reflect, at the voxel level, the previously reported `increasing' or `decreasing' behaviour of TACs within the tumour, which could potentially be linked to intra-tumour aggressiveness heterogeneity. Such map could be of great interest for tumour characterisation as well as for surgery and radiotherapy planning besides conventional static PET imaging.

\section{Acknowledgements}
C. Martens is funded by FRIA grant no.5120417F (F.R.S.-FNRS – Belgian National Fund for Scientific Research). C. Decaestecker is senior research associate with F.R.S.-FNRS. L. Lebrun is funded by Fonds Erasme. The Department of Nuclear Medicine at H\^opital Erasme is supported by Association Vin\c cotte Nuclear (AVN), Fonds Erasme and the Walloon Region (Biowin). The Department of Pathology at H\^opital Erasme is supported by Fonds Erasme and Fonds Yvonne Bo\"el. The authors would like to thank the technical, medical, and scientific staff in charge of PET tracer synthesis and image acquisitions for the needs of this work.

\bibliographystyle{unsrt}
\bibliography{bibliography}

\appendix
\renewcommand{\theequation}{A\arabic{equation}}
\renewcommand{\thefigure}{A\arabic{figure}}
\renewcommand{\thetable}{A\arabic{table}}
\setcounter{equation}{0}
\setcounter{figure}{0}
\setcounter{table}{0} 

\section{Supplementary methods}
\subsection{Patient characteristics}
Patient clinical data, 2016 WHO classification and undergone treatments at imaging time are provided for each lesion in \Cref{table_a1}.

\begin{sidewaystable}[ht!]
\centering
\resizebox{\textwidth}{!}{
\begin{tabular}{r|ccl|lccc|ccc|}
\cline{2-11}
\multicolumn{1}{c|}{} & \multicolumn{3}{c|}{\textbf{Clinical}} & \multicolumn{4}{c|}{\textbf{2016 WHO Classification}} & \multicolumn{3}{c|}{\textbf{Treatments}} \\ \hline
\multicolumn{1}{|c|}{\textbf{Patient}} & \textbf{Age} & \textbf{Sex} & \multicolumn{1}{c|}{\textbf{Location}} & \multicolumn{1}{c}{\textbf{Histology}} & \textbf{Grade} & \textbf{IDH} & \textbf{1p/19q} & \textbf{Surgery} & \textbf{Chemotherapy} & \textbf{Radiotherapy} \\ \hline
\rowcolor[HTML]{C0C0C0} 
\multicolumn{1}{|r|}{\cellcolor[HTML]{C0C0C0}1} & 67 & M & Right post-rolandic & Astrocytoma & III & MT & N/A & Yes & TMZ & RT \\
\multicolumn{1}{|r|}{2} & 53 & F & Right fronto-callosum & Glioblastoma & IV & MT & N/A & Yes & TMZ & RT \\
\rowcolor[HTML]{C0C0C0} 
\multicolumn{1}{|r|}{\cellcolor[HTML]{C0C0C0}3} & 33 & M & Left frontal & Oligodendroglioma & III & MT & CD & Yes & TMZ & RT \\
\multicolumn{1}{|r|}{4} & 56 & F & Right fronto-parietal & Astrocytoma & III & WT & NC & Yes & TMZ & RT, GK \\
\rowcolor[HTML]{C0C0C0} 
\multicolumn{1}{|r|}{\cellcolor[HTML]{C0C0C0}5} & 47 & F & Left frontal & Oligodendroglioma & III & MT & CD & Yes & TMZ & RT \\
\multicolumn{1}{|r|}{6} & 23 & M & Left parietal & Oligodendroglioma & III & MT & CD & Yes & TMZ & RT \\
\rowcolor[HTML]{C0C0C0} 
\multicolumn{1}{|r|}{\cellcolor[HTML]{C0C0C0}7} & 48 & M & Right pre-rolandic & \cellcolor[HTML]{C0C0C0}Oligodendroglioma & II & MT & CD & Yes & TMZ & No \\
\multicolumn{1}{|r|}{8} & 56 & M & Right frontal & Glioblastoma & IV & MT & NC & Yes & No & No \\
\rowcolor[HTML]{C0C0C0} 
\multicolumn{1}{|r|}{\cellcolor[HTML]{C0C0C0}9} & 63 & M & Left temporal & Glioblastoma & IV & WT & NC & Yes & TMZ & RT \\
\multicolumn{1}{|r|}{10} & 34 & F & Thalamo-mesencephalic & Rosette-forming glioneural tumour & I & N/A & N/A & No & No & RT \\
\rowcolor[HTML]{C0C0C0} 
\multicolumn{1}{|r|}{\cellcolor[HTML]{C0C0C0}11} & 36 & F & Right frontal & N/A & N/A & N/A & N/A & No & No & No \\
\multicolumn{1}{|r|}{12} & 52 & F & Left frontal & Glioblastoma & IV & WT & NC & Yes & TMZ & RT, GK \\
\rowcolor[HTML]{C0C0C0} 
\multicolumn{1}{|r|}{\cellcolor[HTML]{C0C0C0}13} & 51 & F & Left parietal & \cellcolor[HTML]{C0C0C0}Oligodendroglioma & III & N/A & CD & Yes & TMZ & RT \\
\multicolumn{1}{|r|}{14} & 58 & M & Right fronto-parietal & Glioblastoma & IV & WT & NC & Yes & TMZ & RT \\
\rowcolor[HTML]{C0C0C0} 
\multicolumn{1}{|r|}{\cellcolor[HTML]{C0C0C0}15} & 56 & F & Left frontal & Astrocytoma & III & MT & NC & Yes & TMZ & RT \\
\multicolumn{1}{|r|}{16} & 50 & M & Left frontal & Oligodendroglioma & III & MT & CD & Yes & TMZ & RT \\
\rowcolor[HTML]{C0C0C0} 
\multicolumn{1}{|r|}{\cellcolor[HTML]{C0C0C0}17} & 61 & M & Right fronto-temporal & Oligodendroglioma & II & MT & CD & Yes & TMZ & No \\
\multicolumn{1}{|r|}{18} & 55 & M & Right frontal & Astrocytoma & III & MT & NC & Yes & TMZ & GK \\
\rowcolor[HTML]{C0C0C0} 
\multicolumn{1}{|r|}{\cellcolor[HTML]{C0C0C0}} & \cellcolor[HTML]{C0C0C0} & \cellcolor[HTML]{C0C0C0} & \cellcolor[HTML]{C0C0C0}Brainstem & N/A & N/A & N/A & N/A & No & \cellcolor[HTML]{C0C0C0} & RT \\
\rowcolor[HTML]{C0C0C0} 
\multicolumn{1}{|r|}{\multirow{-2}{*}{\cellcolor[HTML]{C0C0C0}19}} & \multirow{-2}{*}{\cellcolor[HTML]{C0C0C0}83} & \multirow{-2}{*}{\cellcolor[HTML]{C0C0C0}M} & \cellcolor[HTML]{C0C0C0}Left fronto-insular & N/A & N/A & N/A & N/A & No & \multirow{-2}{*}{\cellcolor[HTML]{C0C0C0}TMZ} & No \\
\multicolumn{1}{|r|}{20} & 58 & M & Right frontal & Glioblastoma & IV & WT & NC & Yes & TMZ & RT, GK \\
\rowcolor[HTML]{C0C0C0} 
\multicolumn{1}{|r|}{\cellcolor[HTML]{C0C0C0}21} & 52 & F & Right parieto-occipital & Glioblastoma & IV & WT & N/A & Yes & TMZ, CCNU & RT \\
\multicolumn{1}{|r|}{22} & 50 & M & Left temporal & Glioblastoma & IV & WT & NC & Yes & TMZ & RT \\
\rowcolor[HTML]{C0C0C0} 
\multicolumn{1}{|r|}{\cellcolor[HTML]{C0C0C0}23} & 70 & F & Left frontal & Glioblastoma & IV & WT & NC & No & No & No \\
\multicolumn{1}{|r|}{} &  &  & Right frontal & Oligodendroglioma & III & MT & CD & Yes &  & RT \\
\multicolumn{1}{|r|}{\multirow{-2}{*}{24}} & \multirow{-2}{*}{35} & \multirow{-2}{*}{F} & Left insular & N/A & N/A & N/A & N/A & No & \multirow{-2}{*}{TMZ} & No \\
\rowcolor[HTML]{C0C0C0} 
\multicolumn{1}{|r|}{\cellcolor[HTML]{C0C0C0}25} & 55 & F & Left temporal & Glioblastoma & IV & WT & NC & No & TMZ & RT \\
\multicolumn{1}{|r|}{26} & 61 & M & Right fronto-temporo-insular & Oligodendroglioma & II & MT & CD & Yes & TMZ & No \\
\rowcolor[HTML]{C0C0C0} 
\multicolumn{1}{|r|}{\cellcolor[HTML]{C0C0C0}27} & 53 & M & Left fronto-insular & \cellcolor[HTML]{C0C0C0}Oligodendroglioma & III & MT & CD & Yes & TMZ & RT \\
\multicolumn{1}{|r|}{28} & 53 & M & Right parietal & Glioblastoma & IV & WT & NC & Yes & TMZ, CCNU & RT \\
\rowcolor[HTML]{C0C0C0} 
\multicolumn{1}{|r|}{\cellcolor[HTML]{C0C0C0}29} & 60 & M & Right frontal & \cellcolor[HTML]{C0C0C0}Glioblastoma & \cellcolor[HTML]{C0C0C0}IV & \cellcolor[HTML]{C0C0C0}WT & NC & Yes & TMZ & RT \\
\multicolumn{1}{|r|}{30} & 76 & F & Left fronto-parietal & Glioblastoma & IV & WT & NC & Yes & TMZ & RT \\
\rowcolor[HTML]{C0C0C0} 
\multicolumn{1}{|r|}{\cellcolor[HTML]{C0C0C0}31} & \cellcolor[HTML]{C0C0C0}42 & \cellcolor[HTML]{C0C0C0}M & \cellcolor[HTML]{C0C0C0}Left frontal & Glioblastoma & IV & WT & \cellcolor[HTML]{C0C0C0}NC & Yes & TMZ & RT \\
\multicolumn{1}{|r|}{} &  &  & Right parieto-occipital & Glioblastoma & IV & WT & NC & Yes &  & RT \\
\multicolumn{1}{|r|}{} &  &  & Right frontal & Glioblastoma & IV & WT & NC & Yes &  & RT \\
\multicolumn{1}{|r|}{\multirow{-3}{*}{32}} & \multirow{-3}{*}{63} & \multirow{-3}{*}{M} & Left frontal & N/A & N/A & N/A & N/A & No & \multirow{-3}{*}{TMZ, CCNU} & No \\
\rowcolor[HTML]{C0C0C0} 
\multicolumn{1}{|r|}{\cellcolor[HTML]{C0C0C0}33} & 56 & M & Right temporo-insular & Astrocytoma & III & MT & NC & \cellcolor[HTML]{C0C0C0}Yes & \cellcolor[HTML]{C0C0C0}TMZ & \cellcolor[HTML]{C0C0C0}RT \\ \hline
\end{tabular}}
\caption{Patient clinical data, location and 2016 WHO classification of each analysed lesion and undergone treatments at imaging time. F: female, M: male, N/A: not available, MT: mutant, WT: wildtype, CD: codeleted, NC: non-codeleted, TMZ: temozolomide, CCNU: lomustin, RT: conventional radiotherapy, GK: gamma knife. Multiple lesions were identified for patients 19, 24 and 32. Patients 1 to 20 and 21 to 33 respectively belong to the training and test set, used for PCA model building and evaluation.}
\label{table_a1}
\end{sidewaystable}

\subsection{Spill-out estimation}
\label{section_a_2}
Spill-out estimation for IDIF correction was performed both analytically and by means of computer simulations. For the sake of simplicity, scanner resolution was considered spatially constant and isotropic, characterised by a full-width at half-maximum (FWHM) of 4.5 mm consistent with our scanner specification. Scanner PSF $g$ was modelled by a 3D isotropic Gaussian with standard deviation $\sigma = \text{FWHM}/\left(2 \sqrt{2 \ln(2)}\right) \approx 1.91$ mm. The petrous segment of the carotid artery from which whole blood input function is extracted in the image was modelled by a cylinder with radius $R=2.4$ mm \cite{kamenskiy2015} and height $H=30.0$ mm, determined experimentally on time-of-flight MR data. For spill-out coefficient calculation, an activity distribution was constructed for such a cylinder centred at the origin, given by;
\begin{equation}
    f(x, y, z) = 
    \begin{cases}
        1, &\text{if } (x, y, z) \in \text{cylinder}\\
        0, &\text{otherwise }
    \end{cases}
    \label{equation_a1} 
\end{equation}
The convolution $c$ of $f$ with the scanner PSF $g$ at point $(x, y, z)$ is then given by:
\begin{align}
    c(x, y, z) &= f(x, y, z) \ast g(x, y, z) 
    \label{equation_a2} \\
    &= \int\limits_{t} \int\limits_{u} \int\limits_{v} f(t, u, v) \, g(x-t, y-u, z-v) \, \mathrm{d}v \, \mathrm{d}u \, \mathrm{d}t
    \label{equation_a3} \\
    &= \frac{1}{\left(2\pi\sigma^2\right)^{\frac{3}{2}}} \int\limits_{t} \int\limits_{u} \int\limits_{v} f(t, u, v) \, e^{-\frac{(x-t)^2+(y-u)^2+(z-v)^2}{2\sigma^2}} \, \mathrm{d}v \, \mathrm{d}u \, \mathrm{d}t
    \label{equation_a4}
\end{align}
To solve \Cref{equation_a4}, the problem was re-expressed in cylindrical coordinates using the following change of variables:
\begin{align}
    t &= r \cos{\theta} 
    \label{equation_a5} \\
    u &= r \sin{\theta} 
    \label{equation_a6} \\
    v &= q
    \label{equation_a7}
\end{align}
leading to:
\begin{equation}
    c(x, y, z) = \frac{1}{\left(2\pi\sigma^2\right)^{\frac{3}{2}}} \int_{0}^{R} \int_{-\pi}^{\pi} \int_{-\frac{H}{2}}^{\frac{H}{2}} r \, e^{- \left(\frac{(x-r\cos\theta)^2 + (y-r\sin\theta)^2 + (z-q)^2}{2\sigma^2}\right)} \, \mathrm{d}q \, \mathrm{d}\theta \, \mathrm{d}r
    \label{equation_a8}
\end{equation}
Since no generic analytical solution exists for \Cref{equation_a8} at every point $(x, y, z)$, we computed the spill-out coefficient at the central point $(0, 0, 0)$ of the cylinder where it is supposed to be maximum, which gives :
\begin{align}
    c(0, 0, 0) &= \frac{1}{\left(2\pi\sigma^2\right)^{\frac{3}{2}}} \int_{0}^{R} \int_{-\pi}^{\pi} \int_{-\frac{H}{2}}^{\frac{H}{2}} r \, e^{- \frac{r^2\cos^2\theta + r^2\sin^2\theta + q^2}{2\sigma^2}} \, \mathrm{d}q \, \mathrm{d}\theta \, \mathrm{d}r
    \label{equation_a9} \\
    &= \frac{1}{\left(2\pi\sigma^2\right)^{\frac{3}{2}}} \int_{0}^{R} \int_{-\pi}^{\pi} \int_{-\frac{H}{2}}^{\frac{H}{2}} r \, e^{- \frac{r^2}{2\sigma^2}} \, e^{- \frac{q^2}{2\sigma^2}} \, \mathrm{d}q \, \mathrm{d}\theta \, \mathrm{d}r
    \label{equation_a10} \\
    &= \frac{1}{\left(2\pi\sigma^2\right)^{\frac{3}{2}}} \frac{\left(\pi\sigma^2\right)^\frac{1}{2}}{2^{\frac{1}{2}}} \left(\erf\left(\frac{\frac{H}{2}}{\sqrt{2}\sigma}\right) - \erf\left(\frac{-\frac{H}{2}}{\sqrt{2}\sigma}\right)\right) \int_{0}^{R} \int_{-\pi}^{\pi} r \, e^{- \frac{r^2}{2\sigma^2}} \, \mathrm{d}\theta \, \mathrm{d}r
    \label{equation_a11} \\
    &= \frac{1}{\left(2\pi\sigma^2\right)^{\frac{3}{2}}} \frac{\left(\pi\sigma^2\right)^\frac{1}{2}}{2^{\frac{1}{2}}} \left(\erf\left(\frac{\frac{H}{2}}{\sqrt{2}\sigma}\right) - \erf\left(\frac{-\frac{H}{2}}{\sqrt{2}\sigma}\right)\right) 2\pi \int_{0}^{R} r \, e^{- \frac{r^2}{2\sigma^2}} \, \mathrm{d}r
    \label{equation_a12} \\
    &= \frac{1}{\left(2\pi\sigma^2\right)^{\frac{3}{2}}} \frac{\left(\pi\sigma^2\right)^\frac{1}{2}}{2^{\frac{1}{2}}} \left(\erf\left(\frac{\frac{H}{2}}{\sqrt{2}\sigma}\right) - \erf\left(\frac{-\frac{H}{2}}{\sqrt{2}\sigma}\right)\right) 2\pi \sigma^2 \left(1-e^{- \frac{R^2}{2\sigma^2}}\right)
    \label{equation_a13} \\
    &= \frac{1}{2}\left(1-e^{- \frac{R^2}{2\sigma^2}}\right)\left(\erf\left(\frac{\frac{H}{2}}{\sqrt{2}\sigma}\right) - \erf\left(\frac{-\frac{H}{2}}{\sqrt{2}\sigma}\right)\right)
    \label{equation_a14}
\end{align}
Provided that an analytical expression can be found for \Cref{equation_a8}, the average spill-out coefficient $c_\text{voxel}$ for a voxel of size 2 mm $\times$ 2 mm $\times$ 2 mm at the centre of the petrous segment of the internal carotid would have been obtained by:
\begin{equation}
    c_\text{voxel} = \frac{1}{8} \int_{-1}^{1} \int_{-1}^{1} \int_{-1}^{1} c(x, y, z) \, \mathrm{d}z \, \mathrm{d}y \, \mathrm{d}x
    \label{equation_a15}
\end{equation}

Alternatively, a numerical approximation of $c_\text{voxel}$ was computed by convolution of a 3D image with voxel size 0.02 mm $\times$ 0.02 mm $\times$ 0.02 mm in the centre of which a cylinder of radius $R$ and height $H$ was drawn with a 3D isotropic Gaussian kernel with standard deviation $\sigma/0.02$ pixels truncated at 4 $\sigma$. An approximated value of $c_\text{voxel}$ was obtained by averaging the 100 $\times$ 100 $\times$ 100 central voxels in the convolved image.

\subsection{2TC model}
\label{section_a_3}
Considering first-order kinetics for the transport processes, we have for the 2TCM:
\begin{align}
\frac{dc_1(t)}{dt} &= K_1 c_p(t) - (k_2+k_3) c_1(t) + k_4 c_2(t)
\label{equation_a16} \\
\frac{dc_2(t)}{dt} &= k_3 c_1(t) - k_4 c_2(t)
\label{equation_a17} \\
c_t(t) &= c_1(t) + c_2(t)
\label{equation_a18}
\end{align}
where $c_p(t)$, $c_1(t)$, $c_2(t)$, and $c_t(t)$ are respectively the plasma, first compartment, second compartment and total tissue tracer activity concentration, $K_1$ and $k_2$ are the rate constants for tracer transport across the blood-brain barrier, and $k_3$ and $k_4$ are the rate constants for trans-membrane cell transport in tissues. Applying unilateral Laplace transform and assuming zero initial concentrations yields:
\begin{align}
C_1(s) &= \frac{K_1 (s+k_4)}{s^2 + (k_2+k_3+k_4) s + k_2k_4} C_p(s)
\label{equation_a19} \\
C_2(s) &= \frac{k_3}{s+k_4} C_1(s) = \frac{K_1 k_3}{s^2 + (k_2+k_3+k_4) s + k_2k_4} C_p(s)
\label{equation_a20} \\
C_t(s) &= \frac{K_1 s + K_1 (k_3+k_4)}{s^2 + (k_2+k_3+k_4) s + k_2k_4} C_p(s)  \quad \Leftrightarrow \quad H_2(s) = \frac{C_t(s)}{C_p(s)} = \frac{K_1 s + K_1 (k_3+k_4)}{s^2 + (k_2+k_3+k_4) s + k_2k_4}
\label{equation_a21}
\end{align}
where $H_2(s)$ is the transfer function of the 2TCM. Using \Cref{equation_4}, making the same assumptions as for the 1TCM regarding \Cref{equation_5} and taking into account carotid-to-voxel delay, the voxel TAC is finally given by the same expression as in \Cref{equation_7} but where $h_1$ is replaced by the Laplace transform $h_2$ of $H_2$.

\begin{figure}[ht!]
\centering
\includegraphics[width=0.9\linewidth]{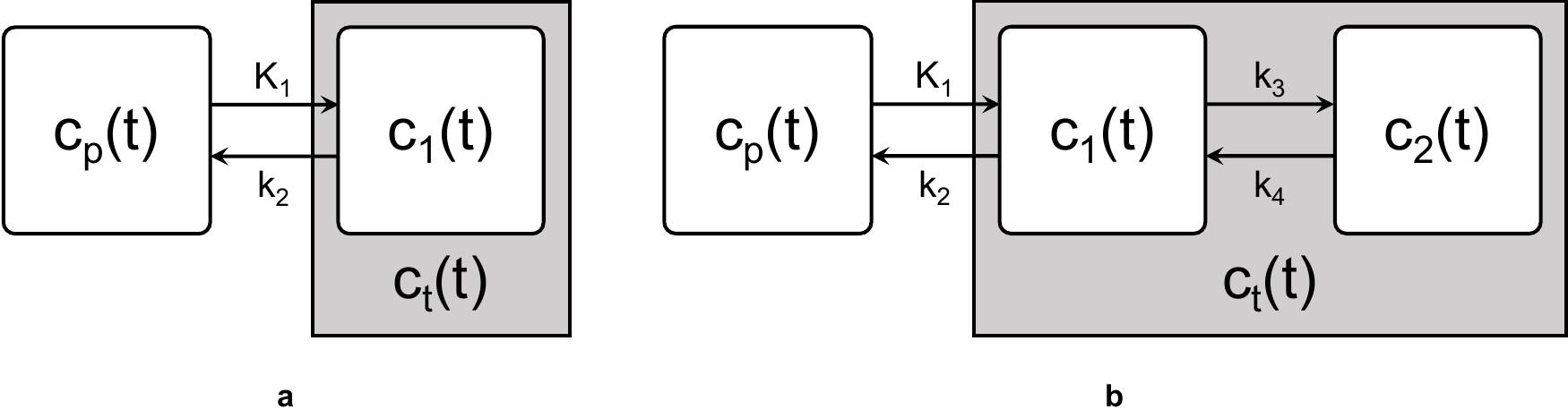}
\caption{Investigated pharmacokinetic models. \textbf{a}: One-tissue compartment model (1TCM) describing the tracer transport from blood to tissues. $c_p(t)$ and $c_1(t)$ are respectively the tracer concentration in the blood and tissue compartment and $K_1$ and $k_2$ are the associated transport rate constants. \textbf{b}: Two-tissue compartment model (2TCM) describing the tracer transport from blood to tissues accross the blood-brain barrier. $c_p(t)$, $c_1(t)$, $c_2(t)$ are respectively the tracer concentration in the blood, extracellular and intracellular compartment. $K_1$ / $k_2$ and $k_3$ / $k_4$ are the rate constants associated with transport across the blood-brain barrier and the cell plasma membrane, respectively.}
\label{figure_a1}
\end{figure}

\subsection{Overlapping frames}
\label{section_a_4}
To quantify the added value of overlapping frames over distinct adjacent frames for the estimation of the 1TCM kinetic parameters, the following numerical experiments were conducted. Feng input function parameters in \Cref{equation_8} were first least-squared fitted for each of the 33 patients' IDIF using the Python \textit{SciPy}'s `optimize' and `signal' modules \cite{virtanen2020}. Frame-averaged blood input functions were then computed for each of the 33 fitted Feng parameter combinations using \Cref{equation_12} and for two distinct framing strategies: (i) 906 overlapping frames of length 20 seconds with a shift of 2 seconds (the framing used in this work) and (ii) 92 non-overlapping adjacent frames of length 20 seconds. For the same two framing strategies, frame-averaged voxel TACs were generated using \Cref{equation_11,equation_12,equation_13} for 10,000 combinations of kinetic parameter values randomly chosen among the whole set of kinetic parameters values previously fitted on the real patients TACs (33 patients, 6,420,534 curves) and for the corresponding Feng parameter values. For each noise level in range 0.0 to 2.0 with step 0.5 and for both framing strategies, synthetic noise was added to the frame-averaged blood input function and to each of the 10,000 frame-averaged voxel TACs using \Cref{equation_14}. Kinetic parameter values were finally estimated from each pair of noisy frame-averaged blood input function and voxel TAC using our least-squares transfer function fitting routine in Python and the relative errors between ground truth and estimated kinetic parameter values were computed.

\section{Supplementary results}
\subsection{Spill-out estimation}
\label{section_b_1}
\begin{figure}[ht!]
\centering
\includegraphics[width=0.6\linewidth]{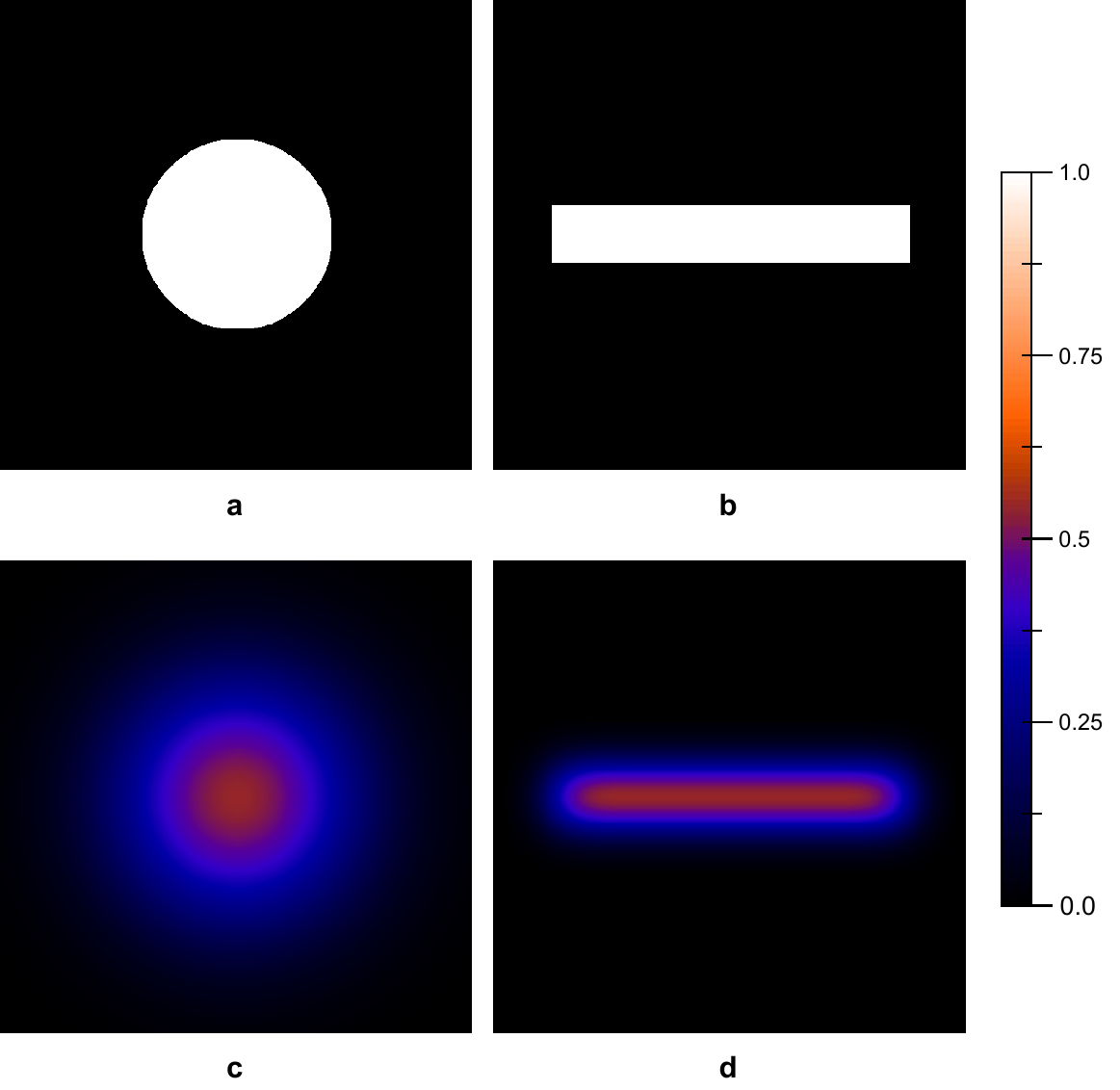}
\caption{Numerical convolution results. Modelled petrous segment of the internal carotid artery filled with unit activity concentration in transverse (\textbf{a}) and axial (\textbf{b}) planes and corresponding simulated activity concentration after convolution with the scanner PSF modelled by an isotropic Gaussian kernel with standard deviation $\sigma=1.91$ mm (\textbf{c}-\textbf{d}).}
\label{figure_a2}
\end{figure}

The evaluation of \Cref{equation_a14} for respective values of $R$ and $H$ of 2.4 mm and 30.0 mm gave a maximum spill-out coefficient $c(0, 0, 0) \approx 0.55$. Numerical convolution results are depicted in \Cref{figure_a2}. The spill-out coefficient value of the central 0.02 mm $\times$ 0.02 mm $\times$ 0.02 mm simulation grid voxel was $0.55$, in accordance with the analytical value. The average spill-out coefficient value within the 100 $\times$ 100 $\times$ 100 central voxels of the simulation grid, corresponding to the central PET image voxel, was 0.51.

\subsection{Overlapping frames}
\label{section_b_2}
\begin{figure}[ht!]
\centering
\includegraphics[width=0.6\linewidth]{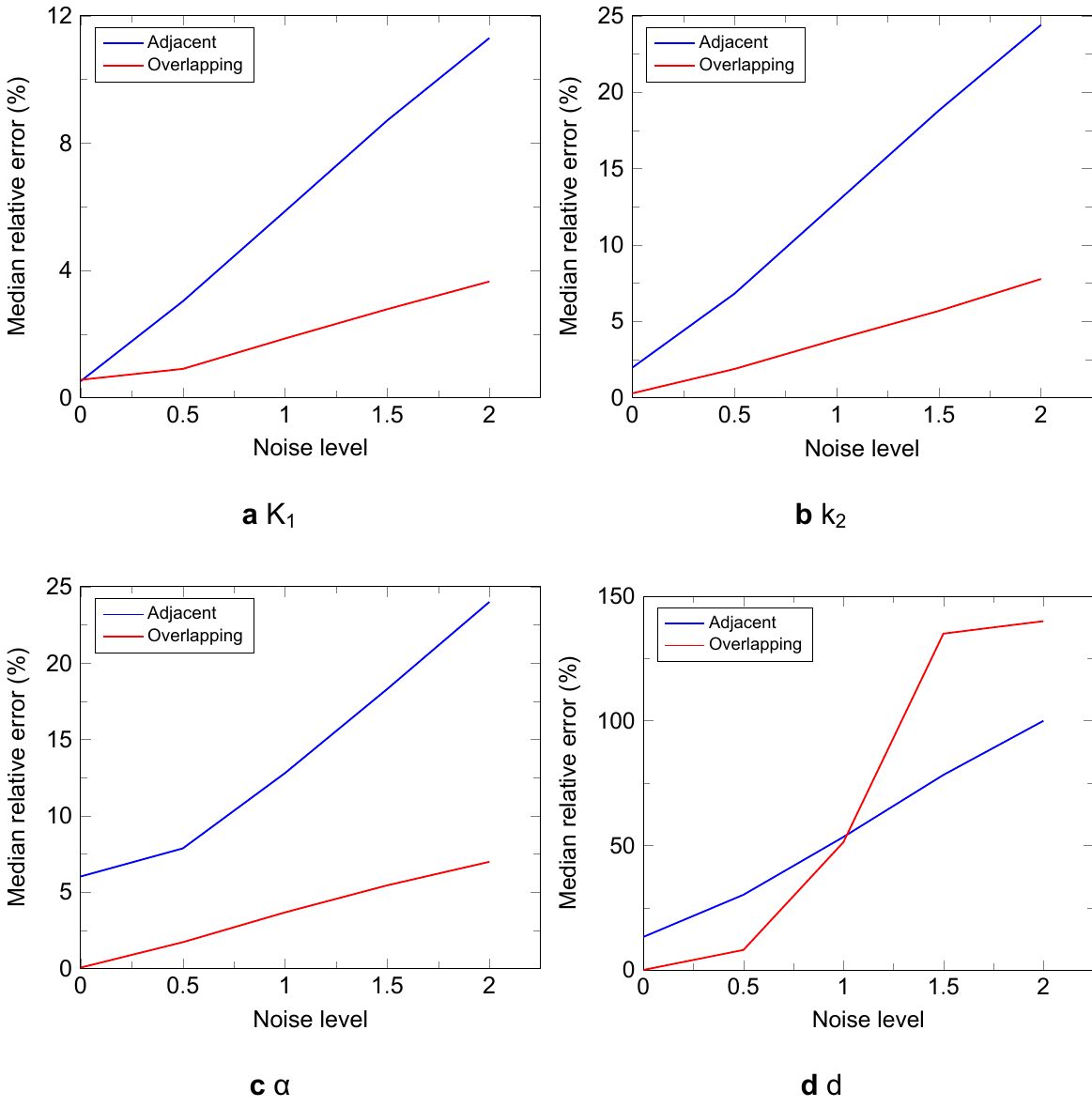}
\caption{Median relative errors on the fitted 1TCM parameter values computed over the 10,000 simulated TACs for each level of noise and both framing strategies (blue curve: adjacent, red curve: overlapping).}
\label{figure_a3}
\end{figure}

Median relative errors on the fitted 1TCM parameter values computed over the 10,000 simulated TACs are depicted in \Cref{figure_a3} for each level of noise and both framing strategies. Except for the delay parameter $d$, overlapping frames systematically provides lower median relative errors. Furthermore, reduction of the fitting error for overlapping compared to adjacent frames increases with the level of noise. Median error decrease for a realistic noise level of 1.5 was of 5.9\%, 13.1\% and 12.9\% for $K_1$, $k_2$ and $\alpha$, respectively. Associated p-values returned by the Wilcoxon's signed-rank test were not considered reliable due to the large number of samples ($n=10,000$), hence Wilcoxon's effect sizes are reported instead in \Cref{table_a2}.

\begin{table}[ht!]
\centering
\begin{tabular}{l|ccccc|}
\cline{2-6}
 & \multicolumn{5}{c|}{\textbf{Noise level}} \\ \cline{2-6} 
 & \textbf{0.0} & \textbf{0.5} & \textbf{1.0} & \textbf{1.5} & \textbf{2.0} \\ \hline
\rowcolor[HTML]{C0C0C0} 
\multicolumn{1}{|l|}{\cellcolor[HTML]{C0C0C0}\textbf{$K_1$}} & 0.12 & 0.72 & 0.70 & 0.70 & 0.70 \\
\multicolumn{1}{|l|}{\textbf{$k_2$}} & 0.87 & 0.74 & 0.71 & 0.71 & 0.70 \\
\rowcolor[HTML]{C0C0C0} 
\multicolumn{1}{|l|}{\cellcolor[HTML]{C0C0C0}\textbf{$\alpha$}} & 0.87 & 0.77 & 0.72 & 0.72 & 0.72 \\
\multicolumn{1}{|l|}{\textbf{$d$}} & 0.87 & 0.70 & -0.10 & -0.67 & -0.59 \\ \hline
\end{tabular}
\caption{Wilcoxon's effect sizes of the difference in relative error between overlapping and adjacent framing strategies for each parameter and each noise level. Negative values correspond to an increase in the absolute error for overlapping frames with regard to adjacent frames.}
\label{table_a2}
\end{table}

\subsection{2TCM parametric maps}
\label{section_b_3}
\begin{figure}[ht!]
\centering
\includegraphics[width=0.9\linewidth]{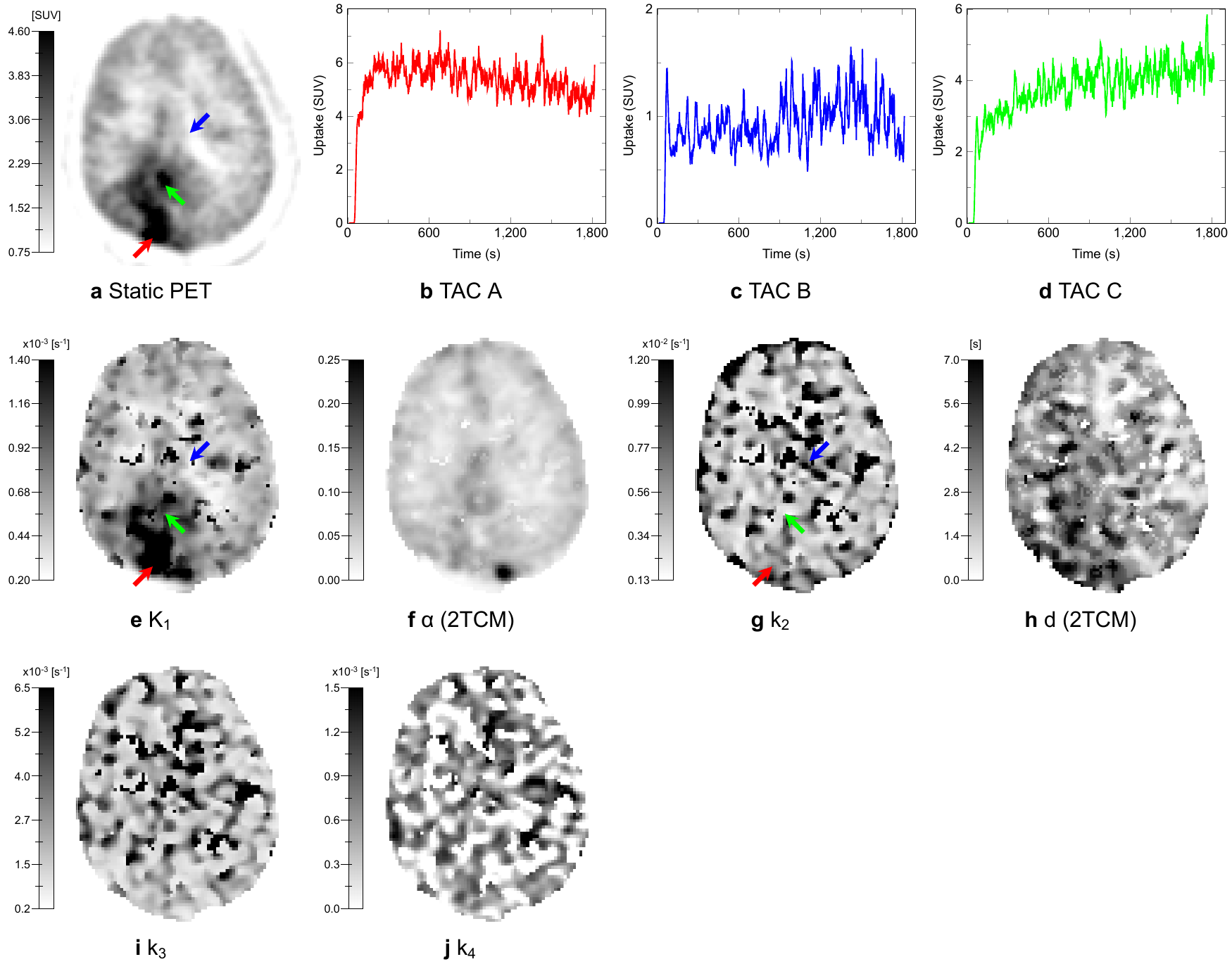}
\caption{2TCM kinetic parameter maps generated from a dynamic [\textsuperscript{11}C]MET PET scan of a glioblastoma test patient (patient 21). \textbf{a}: Static PET (20 to 27 min p.i.). \textbf{b}-\textbf{d}: Smoothed TACs at voxels pointed by the red, blue and green arrows respectively. TAC A, B and C respectively have a `decreasing', `flat', and `increasing' behaviour. \textbf{e}-\textbf{j}: 2TCM kinetic parametric maps. All maps are displayed in inverted greyscale.}
\label{figure_a4}
\end{figure}

2TCM kinetic parameter maps computed from the voxel TACs of a glioblastoma test patient (patient 21) are depicted in \Cref{figure_a4} in inverted greyscale. Static PET image (20 to 27 min p.i.) is depicted in \Cref{figure_a4}\textbf{a}. Curves A, B and C (\Cref{figure_a4}\textbf{b} to \textbf{d}) correspond to the smoothed TACs at voxels pointed by the red, blue, and green arrows respectively. $K_1$ map (\Cref{figure_a4}\textbf{e}) exhibits moderate similarities with static PET image (\Cref{figure_a4}\textbf{a}), PC1 map (Fig. 6\textbf{e}), and $G$ map (Fig. 6\textbf{i}). $\alpha$ (2TCM) map (\Cref{figure_a4}\textbf{f}) exhibits strong similarities with PC2 map (Fig. 6\textbf{f}) and $\alpha$ (1TCM) map (Fig. 6\textbf{j}). $d$ (2TCM) map has similarities with $d$ (1TCM) map (Fig. 6\textbf{l}) and exhibits patterns fairly similar to PC5 map (Fig. 6\textbf{h}) with inverted contrast. $k_2$, $k_3$, and $k_4$ parametric maps (\Cref{figure_a4}\textbf{g}, \textbf{i} and \textbf{j}), on the other hand, do not highlight any distinguishable structures within and outside the BTV. 

\subsection{PET data analysis}
\Cref{table_a3} reports biological tumour volume (BTV), mean and maximum tumour-to-background ratio (TBR) evaluated on the static PET image (20 to 27 min p.i.) as well as tumour contrast evaluated on the static PET image (C\textsubscript{static}) and on the PC1 (C\textsubscript{PC1}) map, given by:
\begin{equation}
\text{C} = \frac{\mid\text{tumour}_\text{mean}-\text{contralateral}_\text{mean}\mid}{\mid\text{tumour}_\text{mean}+\text{contralateral}_\text{mean}\mid}
\label{equation_a22}
\end{equation}
where $\text{tumour}_\text{mean}$ and $\text{contralateral}_\text{mean}$ are respectively the mean value within the BTV and the contralateral spherical ROI (see Methods). Minimum, maximum and mean values of the 1TCM kinetic parameters and of the PC values within the BTV are respectively reported for each lesion in \Cref{table_a4,table_a5}.

\begin{table}[ht!]
\centering
\begin{tabular}{|r|r|r|r|r|r|}
\hline
\multicolumn{1}{|c|}{\textbf{Patient}} & \multicolumn{1}{c|}{\textbf{BTV (mm\textsuperscript{3})}} & \multicolumn{1}{c|}{\textbf{TBR\textsubscript{max}}} & \multicolumn{1}{c|}{\textbf{TBR\textsubscript{mean}}} & \multicolumn{1}{c|}{\textbf{C\textsubscript{static}}} & \multicolumn{1}{c|}{\textbf{C\textsubscript{PC1}}} \\ \hline
\rowcolor[HTML]{C0C0C0} 
1 & 0 & - & - & - & - \\
2 & 32 & 1.67 & 1.60 & 0.23 & 1.25 \\
\rowcolor[HTML]{C0C0C0} 
3 & 664 & 2.46 & 1.90 & 0.31 & 2.99 \\
4 & 2,872 & 2.74 & 1.95 & 0.32 & 1.11 \\
\rowcolor[HTML]{C0C0C0} 
5 & 40 & 1.69 & 1.62 & 0.24 & 1.53 \\
6 & 24 & 1.72 & 1.68 & 0.25 & 3.77 \\
\rowcolor[HTML]{C0C0C0} 
7 & 2,920 & 2.21 & 1.79 & 0.28 & 0.74 \\
8 & 120,808 & 4.80 & 2.35 & 0.40 & 0.79 \\
\rowcolor[HTML]{C0C0C0} 
9 & 1,688 & 2.54 & 1.85 & 0.30 & 0.49 \\
10 & 6,992 & 2.46 & 1.90 & 0.31 & 0.91 \\
\rowcolor[HTML]{C0C0C0} 
11 & 928 & 1.99 & 1.73 & 0.27 & 3.40 \\
12 & 95,744 & 4.82 & 2.22 & 0.38 & 0.64 \\
\rowcolor[HTML]{C0C0C0} 
13 & 0 & - & - & - & - \\
14 & 2,224 & 2.12 & 1.71 & 0.26 & 1.00 \\
\rowcolor[HTML]{C0C0C0} 
15 & 0 & - & - & - & - \\
16 & 2,880 & 2.17 & 1.74 & 0.27 & 1.54 \\
\rowcolor[HTML]{C0C0C0} 
17 & 4,480 & 3.59 & 2.31 & 0.40 & 0.94 \\
18 & 0 & - & - & - & - \\
\rowcolor[HTML]{C0C0C0} 
\cellcolor[HTML]{C0C0C0} & 7,080 & 2.44 & 1.86 & 0.30 & 1.01 \\
\rowcolor[HTML]{C0C0C0} 
\multirow{-2}{*}{\cellcolor[HTML]{C0C0C0}19} & 9,312 & 4.01 & 2.41 & 0.41 & 1.19 \\
20 & 14,800 & 3.26 & 1.94 & 0.32 & 2.95 \\
\rowcolor[HTML]{C0C0C0} 
21 & 113,648 & 4.06 & 2.01 & 0.34 & 0.68 \\
22 & 7,000 & 2.79 & 1.97 & 0.33 & 1.30 \\
\rowcolor[HTML]{C0C0C0} 
23 & 21,976 & 6.84 & 3.12 & 0.51 & 1.15 \\
 & 65,104 & 3.77 & 2.09 & 0.35 & 0.83 \\
\multirow{-2}{*}{24} & 648 & 2.03 & 1.75 & 0.27 & 0.67 \\
\rowcolor[HTML]{C0C0C0} 
25 & 8,728 & 2.84 & 1.90 & 0.31 & 0.67 \\
26 & 4,896 & 3.83 & 2.37 & 0.41 & 1.11 \\
\rowcolor[HTML]{C0C0C0} 
27 & 18,440 & 3.50 & 2.17 & 0.37 & 0.84 \\
28 & 6,528 & 4.42 & 2.35 & 0.40 & 1.08 \\
\rowcolor[HTML]{C0C0C0} 
29 & 3,240 & 2.18 & 1.75 & 0.27 & 2.43 \\
30 & 4,520 & 2.26 & 1.80 & 0.28 & 0.73 \\
\rowcolor[HTML]{C0C0C0} 
31 & 7,288 & 3.08 & 1.91 & 0.31 & 1.01 \\
 & 10,880 & 2.56 & 1.89 & 0.31 & 0.62 \\
 & 21,416 & 4.06 & 2.27 & 0.39 & 0.78 \\
\multirow{-3}{*}{32} & 5,248 & 3.70 & 2.44 & 0.42 & 0.80 \\
\rowcolor[HTML]{C0C0C0} 
33 & 18,680 & 4.00 & 2.11 & 0.36 & 1.31 \\ \hline
\end{tabular}
\caption{PET data analysis results. BTV: biological tumour volume, TBR\textsubscript{max/mean}: maximum/mean tumour-to-background ratio computed on the static PET image (20 to 27 min p.i.), C\textsubscript{static/PC1}: tumour contrast computed on the static PET image/PC1 map.}
\label{table_a3}
\end{table}

\begin{table}[ht!]
\centering
\begin{tabular}{r|rrr|rrr|rrr|rrr|}
\cline{2-13}
\multicolumn{1}{c|}{\textbf{}} & \multicolumn{3}{c|}{\textbf{$G$}} & \multicolumn{3}{c|}{\textbf{$\alpha$}} & \multicolumn{3}{c|}{\textbf{$\tau$ [10\textsuperscript{3} s]}} & \multicolumn{3}{c|}{\textbf{$d$ [s]}} \\ \hline
\multicolumn{1}{|c|}{\textbf{Patient}} & \multicolumn{1}{c}{\textbf{min}} & \multicolumn{1}{c}{\textbf{max}} & \multicolumn{1}{c|}{\textbf{mean}} & \multicolumn{1}{c}{\textbf{min}} & \multicolumn{1}{c}{\textbf{max}} & \multicolumn{1}{c|}{\textbf{mean}} & \multicolumn{1}{c}{\textbf{min}} & \multicolumn{1}{c}{\textbf{max}} & \multicolumn{1}{c|}{\textbf{mean}} & \multicolumn{1}{c}{\textbf{min}} & \multicolumn{1}{c}{\textbf{max}} & \multicolumn{1}{c|}{\textbf{mean}} \\ \hline
\rowcolor[HTML]{C0C0C0} 
\multicolumn{1}{|r|}{\cellcolor[HTML]{C0C0C0}1} & \cellcolor[HTML]{C0C0C0}- & \cellcolor[HTML]{C0C0C0}- & \cellcolor[HTML]{C0C0C0}- & - & - & - & - & - & - & - & - & - \\
\multicolumn{1}{|r|}{2} & 0.62 & 0.67 & 0.64 & 0.04 & 0.05 & 0.05 & 1.48 & 1.64 & 1.56 & 2.45 & 6.37 & 3.52 \\
\rowcolor[HTML]{C0C0C0} 
\multicolumn{1}{|r|}{\cellcolor[HTML]{C0C0C0}3} & 0.31 & 0.48 & 0.39 & 0.04 & 0.09 & 0.05 & \cellcolor[HTML]{C0C0C0}0.58 & \cellcolor[HTML]{C0C0C0}1.12 & \cellcolor[HTML]{C0C0C0}0.86 & 0.34 & 14.40 & 8.89 \\
\multicolumn{1}{|r|}{4} & 0.57 & 0.94 & 0.70 & 0.04 & 0.07 & 0.05 & 1.01 & 1.97 & 1.31 & 1.30 & 14.67 & 6.22 \\
\rowcolor[HTML]{C0C0C0} 
\multicolumn{1}{|r|}{\cellcolor[HTML]{C0C0C0}5} & 0.61 & 0.74 & 0.65 & 0.06 & 0.08 & 0.07 & \cellcolor[HTML]{C0C0C0}1.10 & \cellcolor[HTML]{C0C0C0}1.53 & \cellcolor[HTML]{C0C0C0}1.20 & 3.06 & 8.63 & 4.76 \\
\multicolumn{1}{|r|}{6} & 0.53 & 1.62 & 0.87 & 0.06 & 0.53 & 0.12 & 0.55 & 2.84 & 1.27 & 0.09 & 17.69 & 7.53 \\
\rowcolor[HTML]{C0C0C0} 
\multicolumn{1}{|r|}{\cellcolor[HTML]{C0C0C0}7} & 0.46 & 0.71 & 0.55 & 0.02 & 0.07 & 0.04 & \cellcolor[HTML]{C0C0C0}1.04 & \cellcolor[HTML]{C0C0C0}2.05 & \cellcolor[HTML]{C0C0C0}1.40 & 0.70 & 16.87 & 8.11 \\
\multicolumn{1}{|r|}{8} & 0.65 & 2.30 & 1.12 & 0.06 & 0.22 & 0.12 & 0.77 & 2.17 & 1.16 & 0.69 & 15.94 & 10.60 \\
\rowcolor[HTML]{C0C0C0} 
\multicolumn{1}{|r|}{\cellcolor[HTML]{C0C0C0}9} & 0.79 & 1.20 & 0.95 & 0.06 & 0.12 & 0.08 & \cellcolor[HTML]{C0C0C0}1.16 & \cellcolor[HTML]{C0C0C0}1.79 & \cellcolor[HTML]{C0C0C0}1.45 & 2.69 & 14.56 & 10.74 \\
\multicolumn{1}{|r|}{10} & 0.54 & 0.89 & 0.70 & 0.06 & 0.28 & 0.09 & 0.76 & 1.29 & 0.96 & 4.09 & 14.60 & 8.97 \\
\rowcolor[HTML]{C0C0C0} 
\multicolumn{1}{|r|}{\cellcolor[HTML]{C0C0C0}11} & 0.47 & 0.58 & 0.52 & 0.02 & 0.09 & 0.04 & \cellcolor[HTML]{C0C0C0}0.82 & \cellcolor[HTML]{C0C0C0}1.71 & \cellcolor[HTML]{C0C0C0}1.17 & 2.54 & 14.81 & 6.45 \\
\multicolumn{1}{|r|}{12} & 0.68 & 2.77 & 1.17 & 0.05 & 0.29 & 0.11 & 0.66 & 1.72 & 1.14 & 0.57 & 16.83 & 9.12 \\
\rowcolor[HTML]{C0C0C0} 
\multicolumn{1}{|r|}{\cellcolor[HTML]{C0C0C0}13} & \cellcolor[HTML]{C0C0C0}- & \cellcolor[HTML]{C0C0C0}- & \cellcolor[HTML]{C0C0C0}- & - & - & - & - & - & - & - & - & - \\
\multicolumn{1}{|r|}{14} & 0.41 & 0.70 & 0.50 & 0.03 & 0.08 & 0.05 & 1.16 & 2.97 & 1.60 & 1.55 & 14.35 & 7.39 \\
\rowcolor[HTML]{C0C0C0} 
\multicolumn{1}{|r|}{\cellcolor[HTML]{C0C0C0}15} & \cellcolor[HTML]{C0C0C0}- & \cellcolor[HTML]{C0C0C0}- & \cellcolor[HTML]{C0C0C0}- & - & - & - & - & - & - & - & - & - \\
\multicolumn{1}{|r|}{16} & 0.46 & 0.66 & 0.55 & 0.04 & 0.08 & 0.06 & 0.83 & 1.92 & 1.13 & 2.27 & 17.53 & 9.00 \\
\rowcolor[HTML]{C0C0C0} 
\multicolumn{1}{|r|}{\cellcolor[HTML]{C0C0C0}17} & 0.45 & 1.18 & 0.71 & 0.02 & 0.13 & 0.06 & \cellcolor[HTML]{C0C0C0}0.89 & \cellcolor[HTML]{C0C0C0}1.90 & \cellcolor[HTML]{C0C0C0}1.23 & 0.67 & 14.60 & 7.23 \\
\multicolumn{1}{|r|}{18} & - & - & - & - & - & - & - & - & - & - & - & - \\
\rowcolor[HTML]{C0C0C0} 
\multicolumn{1}{|r|}{\cellcolor[HTML]{C0C0C0}} & 0.49 & 0.83 & 0.63 & 0.03 & 0.15 & 0.06 & 0.98 & 1.91 & 1.27 & 2.58 & 14.59 & 7.93 \\
\rowcolor[HTML]{C0C0C0} 
\multicolumn{1}{|r|}{\multirow{-2}{*}{\cellcolor[HTML]{C0C0C0}19}} & 0.51 & 1.15 & 0.78 & 0.03 & 0.09 & 0.06 & \cellcolor[HTML]{C0C0C0}0.96 & \cellcolor[HTML]{C0C0C0}2.66 & \cellcolor[HTML]{C0C0C0}1.30 & 2.36 & 12.14 & 6.99 \\
\multicolumn{1}{|r|}{20} & 0.49 & 1.13 & 0.71 & 0.03 & 0.11 & 0.06 & 0.81 & 2.47 & 1.61 & 1.16 & 14.77 & 7.44 \\
\rowcolor[HTML]{C0C0C0} 
\multicolumn{1}{|r|}{\cellcolor[HTML]{C0C0C0}21} & 0.43 & 0.45 & 0.44 & 0.04 & 0.05 & 0.05 & \cellcolor[HTML]{C0C0C0}1.04 & \cellcolor[HTML]{C0C0C0}1.14 & \cellcolor[HTML]{C0C0C0}1.08 & 6.25 & 8.11 & 6.92 \\
\multicolumn{1}{|r|}{22} & 0.72 & 1.45 & 1.06 & 0.06 & 0.18 & 0.12 & 0.80 & 1.94 & 1.35 & 0.50 & 15.16 & 8.68 \\
\rowcolor[HTML]{C0C0C0} 
\multicolumn{1}{|r|}{\cellcolor[HTML]{C0C0C0}23} & 0.51 & 3.03 & 1.19 & 0.06 & 0.28 & 0.16 & \cellcolor[HTML]{C0C0C0}0.46 & \cellcolor[HTML]{C0C0C0}0.96 & \cellcolor[HTML]{C0C0C0}0.79 & 2.20 & 12.19 & 6.74 \\
\multicolumn{1}{|r|}{} & 0.42 & 1.22 & 0.63 & 0.05 & 0.26 & 0.12 & 0.47 & 1.21 & 0.70 & 0.64 & 16.84 & 9.64 \\
\multicolumn{1}{|r|}{\multirow{-2}{*}{24}} & 0.49 & 0.61 & 0.55 & 0.04 & 0.08 & 0.06 & 0.95 & 1.16 & 1.04 & 2.31 & 16.42 & 7.70 \\
\rowcolor[HTML]{C0C0C0} 
\multicolumn{1}{|r|}{\cellcolor[HTML]{C0C0C0}25} & 0.68 & 1.29 & 0.88 & 0.05 & 0.11 & 0.08 & \cellcolor[HTML]{C0C0C0}0.88 & \cellcolor[HTML]{C0C0C0}1.81 & \cellcolor[HTML]{C0C0C0}1.29 & 4.08 & 15.09 & 11.71 \\
\multicolumn{1}{|r|}{26} & 0.40 & 1.13 & 0.69 & 0.02 & 0.13 & 0.05 & 0.98 & 1.89 & 1.32 & 0.10 & 14.28 & 5.78 \\
\rowcolor[HTML]{C0C0C0} 
\multicolumn{1}{|r|}{\cellcolor[HTML]{C0C0C0}27} & 0.47 & 1.21 & 0.73 & 0.02 & 0.07 & 0.04 & \cellcolor[HTML]{C0C0C0}1.08 & \cellcolor[HTML]{C0C0C0}2.62 & \cellcolor[HTML]{C0C0C0}1.51 & 0.60 & 14.61 & 6.93 \\
\multicolumn{1}{|r|}{28} & 0.60 & 1.55 & 0.95 & 0.04 & 0.23 & 0.09 & 0.80 & 2.35 & 1.32 & 2.02 & 14.43 & 7.62 \\
\rowcolor[HTML]{C0C0C0} 
\multicolumn{1}{|r|}{\cellcolor[HTML]{C0C0C0}29} & 0.41 & 0.55 & 0.48 & 0.02 & 0.05 & 0.04 & \cellcolor[HTML]{C0C0C0}0.94 & \cellcolor[HTML]{C0C0C0}1.78 & \cellcolor[HTML]{C0C0C0}1.21 & 0.57 & 14.71 & 6.40 \\
\multicolumn{1}{|r|}{30} & 0.44 & 0.63 & 0.53 & 0.05 & 0.09 & 0.07 & 0.65 & 1.22 & 0.88 & 2.38 & 14.44 & 7.64 \\
\rowcolor[HTML]{C0C0C0} 
\multicolumn{1}{|r|}{\cellcolor[HTML]{C0C0C0}31} & 0.44 & 0.80 & 0.59 & 0.04 & 0.14 & 0.08 & \cellcolor[HTML]{C0C0C0}0.67 & \cellcolor[HTML]{C0C0C0}2.16 & \cellcolor[HTML]{C0C0C0}1.07 & 2.12 & 15.77 & 13.43 \\
\multicolumn{1}{|r|}{} & 0.59 & 1.08 & 0.78 & 0.04 & 0.10 & 0.07 & 1.08 & 1.84 & 1.37 & 6.25 & 17.75 & 13.94 \\
\multicolumn{1}{|r|}{} & 0.53 & 1.76 & 0.86 & 0.03 & 0.20 & 0.10 & 0.97 & 2.31 & 1.30 & 4.43 & 17.38 & 12.60 \\
\multicolumn{1}{|r|}{\multirow{-3}{*}{32}} & 0.55 & 1.46 & 0.90 & 0.05 & 0.16 & 0.11 & 0.93 & 1.59 & 1.20 & 4.10 & 17.06 & 11.45 \\
\rowcolor[HTML]{C0C0C0} 
\multicolumn{1}{|r|}{\cellcolor[HTML]{C0C0C0}33} & 0.40 & 1.17 & 0.71 & 0.04 & 0.22 & 0.08 & \cellcolor[HTML]{C0C0C0}0.64 & \cellcolor[HTML]{C0C0C0}2.80 & \cellcolor[HTML]{C0C0C0}1.50 & 0.56 & 15.01 & 8.11 \\ \hline
\end{tabular}
\caption{Minimum, maximum and mean 1TCM kinetic parameters values computed for each lesion within the BTV.}
\label{table_a4}
\end{table}

\begin{table}[ht!]
\centering
\begin{tabular}{r|rrr|rrr|rrr|rrr|}
\cline{2-13}
\multicolumn{1}{c|}{\textbf{}} & \multicolumn{3}{c|}{\textbf{PC1}} & \multicolumn{3}{c|}{\textbf{PC2}} & \multicolumn{3}{c|}{\textbf{PC3}} & \multicolumn{3}{c|}{\textbf{PC5}} \\ \hline
\multicolumn{1}{|c|}{\textbf{Patient}} & \multicolumn{1}{c}{\textbf{min}} & \multicolumn{1}{c}{\textbf{max}} & \multicolumn{1}{c|}{\textbf{mean}} & \multicolumn{1}{c}{\textbf{min}} & \multicolumn{1}{c}{\textbf{max}} & \multicolumn{1}{c|}{\textbf{mean}} & \multicolumn{1}{c}{\textbf{min}} & \multicolumn{1}{c}{\textbf{max}} & \multicolumn{1}{c|}{\textbf{mean}} & \multicolumn{1}{c}{\textbf{min}} & \multicolumn{1}{c}{\textbf{max}} & \multicolumn{1}{c|}{\textbf{mean}} \\ \hline
\rowcolor[HTML]{C0C0C0} 
\multicolumn{1}{|r|}{\cellcolor[HTML]{C0C0C0}1} & - & - & - & - & - & - & - & - & - & - & - & - \\
\multicolumn{1}{|r|}{2} & 14.28 & 16.96 & 15.71 & -5.36 & -3.91 & -4.69 & 0.33 & 1.39 & 0.81 & -0.32 & 0.16 & -0.11 \\
\rowcolor[HTML]{C0C0C0} 
\multicolumn{1}{|r|}{\cellcolor[HTML]{C0C0C0}3} & 4.27 & 28.78 & 16.24 & -5.27 & 1.37 & -3.17 & \cellcolor[HTML]{C0C0C0}-3.88 & \cellcolor[HTML]{C0C0C0}1.23 & \cellcolor[HTML]{C0C0C0}-1.00 & -1.44 & 1.25 & 0.04 \\
\multicolumn{1}{|r|}{4} & 14.33 & 43.25 & 27.29 & -7.01 & -1.70 & -4.39 & -1.27 & 3.13 & 1.10 & -1.97 & 1.24 & -0.25 \\
\rowcolor[HTML]{C0C0C0} 
\multicolumn{1}{|r|}{\cellcolor[HTML]{C0C0C0}5} & 14.11 & 16.66 & 15.33 & -4.75 & -2.13 & -3.04 & \cellcolor[HTML]{C0C0C0}-0.55 & \cellcolor[HTML]{C0C0C0}1.16 & \cellcolor[HTML]{C0C0C0}-0.07 & -0.59 & 0.79 & 0.37 \\
\multicolumn{1}{|r|}{6} & 31.29 & 158.52 & 60.37 & -9.53 & 40.13 & -3.42 & -13.74 & 7.23 & 0.79 & -2.05 & 6.34 & 0.42 \\
\rowcolor[HTML]{C0C0C0} 
\multicolumn{1}{|r|}{\cellcolor[HTML]{C0C0C0}7} & 20.15 & 47.51 & 35.25 & -6.44 & 2.69 & -3.22 & \cellcolor[HTML]{C0C0C0}-4.21 & \cellcolor[HTML]{C0C0C0}1.66 & \cellcolor[HTML]{C0C0C0}-1.02 & -1.67 & 1.58 & 0.24 \\
\multicolumn{1}{|r|}{8} & 28.37 & 170.40 & 71.41 & -10.45 & 11.80 & -3.25 & -11.69 & 5.26 & -1.16 & -3.85 & 3.76 & -0.72 \\
\rowcolor[HTML]{C0C0C0} 
\multicolumn{1}{|r|}{\cellcolor[HTML]{C0C0C0}9} & 41.57 & 83.09 & 58.68 & -11.16 & -5.85 & -7.83 & \cellcolor[HTML]{C0C0C0}-1.02 & \cellcolor[HTML]{C0C0C0}4.11 & \cellcolor[HTML]{C0C0C0}2.11 & -2.29 & 1.08 & -0.48 \\
\multicolumn{1}{|r|}{10} & 22.72 & 55.64 & 36.90 & -6.26 & 17.13 & -2.05 & -4.63 & 2.04 & -1.09 & -0.83 & 1.62 & 0.24 \\
\rowcolor[HTML]{C0C0C0} 
\multicolumn{1}{|r|}{\cellcolor[HTML]{C0C0C0}11} & 0.52 & 9.10 & 5.84 & -4.09 & 2.80 & -1.58 & \cellcolor[HTML]{C0C0C0}-2.46 & \cellcolor[HTML]{C0C0C0}3.27 & \cellcolor[HTML]{C0C0C0}0.30 & -2.12 & 0.70 & -0.79 \\
\multicolumn{1}{|r|}{12} & 42.19 & 212.93 & 88.56 & -14.81 & 9.22 & -4.26 & -16.05 & 5.26 & -2.21 & -2.72 & 3.54 & 0.34 \\
\rowcolor[HTML]{C0C0C0} 
\multicolumn{1}{|r|}{\cellcolor[HTML]{C0C0C0}13} & - & - & - & - & - & - & - & - & - & - & - & - \\
\multicolumn{1}{|r|}{14} & 14.28 & 16.96 & 15.71 & -5.36 & -3.91 & -4.69 & 0.33 & 1.39 & 0.81 & -0.32 & 0.16 & -0.11 \\
\rowcolor[HTML]{C0C0C0} 
\multicolumn{1}{|r|}{\cellcolor[HTML]{C0C0C0}15} & - & - & - & - & - & - & - & - & - & - & - & - \\
\multicolumn{1}{|r|}{16} & 14.28 & 16.96 & 15.71 & -5.36 & -3.91 & -4.69 & 0.33 & 1.39 & 0.81 & -0.32 & 0.16 & -0.11 \\
\rowcolor[HTML]{C0C0C0} 
\multicolumn{1}{|r|}{\cellcolor[HTML]{C0C0C0}17} & 16.43 & 99.84 & 48.61 & -10.55 & 8.43 & -3.36 & \cellcolor[HTML]{C0C0C0}-5.43 & \cellcolor[HTML]{C0C0C0}5.80 & \cellcolor[HTML]{C0C0C0}-0.87 & -2.25 & 1.92 & -0.52 \\
\multicolumn{1}{|r|}{18} & - & - & - & - & - & - & - & - & - & - & - & - \\
\rowcolor[HTML]{C0C0C0} 
\multicolumn{1}{|r|}{\cellcolor[HTML]{C0C0C0}} & 16.15 & 45.37 & 29.38 & -4.35 & 10.82 & -1.09 & -3.89 & 4.96 & -1.37 & -1.55 & 3.26 & -0.30 \\
\rowcolor[HTML]{C0C0C0} 
\multicolumn{1}{|r|}{\multirow{-2}{*}{\cellcolor[HTML]{C0C0C0}19}} & 12.91 & 81.32 & 43.66 & -5.98 & 2.70 & -1.86 & \cellcolor[HTML]{C0C0C0}-5.30 & \cellcolor[HTML]{C0C0C0}3.29 & \cellcolor[HTML]{C0C0C0}-1.32 & -1.55 & 2.08 & -0.10 \\
\multicolumn{1}{|r|}{20} & 2.93 & 47.21 & 17.24 & -5.62 & 3.56 & -2.18 & -5.51 & 2.85 & 0.25 & -1.74 & 1.98 & -0.22 \\
\rowcolor[HTML]{C0C0C0} 
\multicolumn{1}{|r|}{\cellcolor[HTML]{C0C0C0}21} & 4.87 & 5.68 & 5.31 & -2.11 & -1.67 & -1.90 & \cellcolor[HTML]{C0C0C0}-1.54 & \cellcolor[HTML]{C0C0C0}-0.98 & \cellcolor[HTML]{C0C0C0}-1.34 & -0.15 & 0.17 & 0.05 \\
\multicolumn{1}{|r|}{22} & 10.63 & 43.98 & 27.54 & -4.97 & 5.63 & -1.48 & -5.64 & 3.19 & -0.60 & -1.25 & 3.54 & 0.08 \\
\rowcolor[HTML]{C0C0C0} 
\multicolumn{1}{|r|}{\cellcolor[HTML]{C0C0C0}23} & 11.16 & 176.62 & 64.48 & -12.61 & 5.95 & -4.10 & \cellcolor[HTML]{C0C0C0}-9.51 & \cellcolor[HTML]{C0C0C0}-0.18 & \cellcolor[HTML]{C0C0C0}-3.18 & -6.36 & -0.13 & -2.37 \\
\multicolumn{1}{|r|}{} & 24.51 & 126.14 & 53.75 & -6.08 & 13.17 & 1.03 & -18.79 & 1.11 & -6.77 & -1.02 & 4.55 & 1.12 \\
\multicolumn{1}{|r|}{\multirow{-2}{*}{24}} & 23.93 & 35.83 & 29.72 & -7.02 & -2.86 & -5.03 & -1.16 & 1.05 & -0.03 & -0.72 & 0.62 & -0.23 \\
\rowcolor[HTML]{C0C0C0} 
\multicolumn{1}{|r|}{\cellcolor[HTML]{C0C0C0}25} & 29.09 & 84.23 & 51.18 & -6.44 & 0.59 & -3.91 & \cellcolor[HTML]{C0C0C0}-6.31 & \cellcolor[HTML]{C0C0C0}2.70 & \cellcolor[HTML]{C0C0C0}-1.23 & -1.49 & 1.58 & -0.06 \\
\multicolumn{1}{|r|}{26} & 10.54 & 78.94 & 38.84 & -9.88 & 9.33 & -3.91 & -2.66 & 4.21 & 1.12 & -2.06 & 5.20 & -0.05 \\
\rowcolor[HTML]{C0C0C0} 
\multicolumn{1}{|r|}{\cellcolor[HTML]{C0C0C0}27} & 22.11 & 101.12 & 51.98 & -9.91 & 0.55 & -4.96 & \cellcolor[HTML]{C0C0C0}-6.23 & \cellcolor[HTML]{C0C0C0}3.31 & \cellcolor[HTML]{C0C0C0}-1.43 & -1.45 & 1.94 & 0.14 \\
\multicolumn{1}{|r|}{28} & 12.85 & 99.81 & 46.08 & -4.82 & 22.30 & 0.41 & -7.44 & 1.85 & -2.99 & -1.38 & 2.59 & 0.38 \\
\rowcolor[HTML]{C0C0C0} 
\multicolumn{1}{|r|}{\cellcolor[HTML]{C0C0C0}29} & 7.03 & 25.60 & 16.39 & -4.05 & 0.21 & -2.06 & \cellcolor[HTML]{C0C0C0}-4.56 & \cellcolor[HTML]{C0C0C0}0.76 & \cellcolor[HTML]{C0C0C0}-2.08 & -0.97 & 1.51 & 0.19 \\
\multicolumn{1}{|r|}{30} & 25.53 & 57.29 & 39.09 & -5.05 & 0.10 & -2.52 & -6.53 & 1.57 & -1.99 & -0.36 & 1.93 & 0.56 \\
\rowcolor[HTML]{C0C0C0} 
\multicolumn{1}{|r|}{\cellcolor[HTML]{C0C0C0}31} & 12.26 & 58.58 & 33.05 & -4.99 & 6.12 & -0.62 & \cellcolor[HTML]{C0C0C0}-12.30 & \cellcolor[HTML]{C0C0C0}2.52 & \cellcolor[HTML]{C0C0C0}-3.92 & -1.01 & 4.51 & 0.64 \\
\multicolumn{1}{|r|}{} & 29.90 & 75.90 & 51.22 & -10.82 & -2.89 & -6.93 & -3.31 & 2.67 & 0.02 & -1.77 & 0.63 & -0.55 \\
\multicolumn{1}{|r|}{} & 25.09 & 137.01 & 63.97 & -10.30 & 2.77 & -4.88 & -7.45 & 3.76 & -1.21 & -2.20 & 1.74 & -0.21 \\
\multicolumn{1}{|r|}{\multirow{-3}{*}{32}} & 26.66 & 119.75 & 71.06 & -8.28 & 3.39 & -3.50 & -5.64 & 1.21 & -1.90 & -1.12 & 3.34 & 0.67 \\
\rowcolor[HTML]{C0C0C0} 
\multicolumn{1}{|r|}{\cellcolor[HTML]{C0C0C0}33} & 9.00 & 75.38 & 30.90 & -6.02 & 16.17 & -1.39 & \cellcolor[HTML]{C0C0C0}-10.91 & \cellcolor[HTML]{C0C0C0}4.30 & \cellcolor[HTML]{C0C0C0}0.00 & -2.54 & 4.81 & -0.08 \\ \hline
\end{tabular}
\caption{Minimum, maximum and mean PC values computed for each lesion within the BTV.}
\label{table_a5}
\end{table}

\section{Supplementary discussion}
We showed that reconstruction of a larger number of overlapping frames compared to a restricted number of adjacent frames of the same length benefits the estimation of the 1TCM kinetic parameters increasingly with the noise level, except for the blood delay $d$. Possible explanations for this gain are that: (i) overlapping frames provide a better sampling of the arterial peak for both blood and tissue TACs and (ii) increased number of correlated data points helps the fitting algorithm to overcome noise. The latter remark does however not seem to apply for the estimation of the blood delay $d$, which basically consists in an arterial peak alignment process, negatively impacted by an increased number of early noisy TAC samples. Nevertheless, blood delay $d$ is rarely considered as a variable of interest in pharmacokinetic modelling but is rather used to provide more flexibility to the fitting algorithm. The overlapping frames approach proposed has however two major drawbacks, that are (i) longer reconstruction times and (ii) larger memory usage for storage (Hard Disk Drive) and processing (Random Access Memory). It should finally be noted that the proposed overlapping framing may not be optimal for kinetic parameter values estimation. Determining the best dynamic PET framing strategy for pharmacokinetic modelling would be of interest but is out of the scope of this paper.

We proposed to average TACs of the 10 brightest voxels within the petrous segment of both internal carotid arteries on an early PET frame for blood input function extraction (see Methods). It turned out in practice that these voxels are almost systematically located along the central line of the segment, which is consistent with our simulations since central line has the largest theoretical spill-out coefficient value. Whereas this method has the great advantage of not requiring invasive arterial sampling, we showed hereabove that activity at these voxels is systematically underestimated by a factor near 2 due to partial volume effects resulting from the limited resolution of the scanner and the restricted diameter of the internal carotid arteries. We proposed to address this issue by scaling the extracted blood input functions by a factor $1/0.51 \approx 1.96$. This correction remains however limited since geometry of the carotid arteries varies among patients \cite{kamenskiy2015}. Additionally, investigating the effects of these variations on the kinetic parameters estimation would be of interest but are out of the scope of this work. Spill-in effects, in contrast, are hardly estimated since all TACs within the reconstructed volume theoretically impacts the input function and are still a challenging open problem in PET imaging \cite{arkele2020}.

\end{document}